\documentclass[preprint, endfloats*, amsmath, floatfix, prmaterials, superscriptaddress]{revtex4-2}
\usepackage{graphicx}
\usepackage{hyperref}
\usepackage{xcolor}
\usepackage{subcaption}
\usepackage{multirow}

\begin{document}

\title{Interplay between magnetism and short-range order in {medium- and} high-entropy alloys: CrCoNi, CrFeCoNi, and CrMnFeCoNi}

\author{Christopher D. Woodgate}
\affiliation{Department of Physics, University of Warwick, Coventry, CV4 7AL, United Kingdom}
\author{Daniel Hedlund}
\affiliation{Department of Chemical Engineering, Northeastern University, Boston, MA 02115, USA}
\author{L. H. Lewis}
\affiliation{Department of Chemical Engineering, Northeastern University, Boston, MA 02115, USA}
\affiliation{Department of Mechanical and Industrial Engineering, Northeastern University, Boston, MA 02115, USA}
\author{Julie B. Staunton}
\affiliation{Department of Physics, University of Warwick, Coventry, CV4 7AL, United Kingdom}

\begin{abstract}
The impact of magnetism on predicted atomic short-range order in {three medium- and} high-entropy alloys is studied using a first-principles, all-electron, Landau-type linear response theory, coupled with lattice-based atomistic modelling. 
We perform two sets of linear-response calculations: one in which the paramagnetic state is modelled within the disordered local moment picture, and one in which systems are modelled in a magnetically ordered state{, which is ferrimagnetic for the alloys considered in this work}. 
We show that the treatment of magnetism can have significant impact both on the predicted temperature of atomic ordering and also the nature of atomic order itself.
In CrCoNi, we find that the nature of atomic order changes from being $\mathrm{L}1_2$-like when modelled in the paramagnetic state to MoPt$_2$-like when modelled assuming the system has magnetically ordered. In CrFeCoNi, atomic correlations between Fe and the other elements present are dramatically strengthened when we switch from treating the system as magnetically disordered to magnetically ordered.
Our results show it is necessary to consider the magnetic state when modelling multicomponent alloys containing mid- to late-$3d$ elements.
Further, we suggest that there may be high-entropy alloy compositions containing $3d$ transition metals that will exhibit specific atomic short-range order when thermally treated in an applied magnetic field.
{ This has the potential to provide a route for tuning physical and mechanical properties in this class of materials.}
\end{abstract}

\maketitle

\section{Introduction}
\label{sec:introduction}

High-entropy alloys, formed by combining four or more elements in near-equal ratios, are of considerable current interest in the field of materials science due to their superior physical and mechanical properties for a range of applications. 
These properties include exceptional hardness, high tensile strength, and excellent resistance to wear, irradiation, and fracture\cite{zhang_microstructures_2014, pickering_high-entropy_2016, gao_high-entropy_2016, miracle_critical_2017, george_high-entropy_2019}.
The space of potential alloy compositions is vast and, therefore, considerable effort has gone into modelling these materials computationally to guide experiments towards suitable compositions, using a variety of approaches. These include first-principles calculations of materials' electronic structure, approaches based on effective medium theories such as the coherent potential approximation (CPA), machine-learned interatomic potentials, and CALPHAD\cite{singh_atomic_2015, troparevsky_criteria_2015, liu_thermodynamic_2016, gao_thermodynamics_2017, fernandez-caballero_short-range_2017, widom_modeling_2018, ikeda_ab_2019, kostiuchenko_impact_2019, ferrari_frontiers_2020, wang_thermodynamic_2021}.

The single-phase solid solutions experimentally realized in this class of materials are stabilised by the large contribution from the configurational entropy to the free energy of the system\cite{yeh_nanostructured_2004}. Despite no reported experimental observations of atomic long-range order in these alloys, atomic short-range order (ASRO) has been both predicted computationally and observed experimentally \cite{zhang_local_2017, zhang_short-range_2020, li_evolution_2023}. 
Atomic short-range order is known to impact a wide range of materials properties \cite{zhang_short-range_2020, wu_short-range_2021, li_evolution_2023, ziehl_detection_2023} and it is therefore important to understand the nature of ASRO and identify the temperatures at which it becomes significant to help guide experiment. Moreover, with the intrinsic link to electronic structure, the influence of magnetic order on ASRO can be profound.

One experimental technique that can alter the magnetic state of a material, and in turn influence its spin polarisation, is magnetic field annealing. This technique is reported to have an effect on a wide range of magnetic as well as non-magnetic materials.
In the late 1800s and early 1900s it was shown that annealing steel under applied static or alternating magnetic field conditions can influence its magnetic properties~\cite{guthe_influence_1897, pender_annealing_1913, fuji_hiromichi_effects_2009}. 
Further, the application of a magnetic field during thermal processing of the technologically important alnico magnets can induce magnetic anisotropy, donating improved permanent magnetic performance~\cite{mccurrie_chapter_1982}.
Yet another class of materials affected by magnetic field annealing is soft/ultrasoft nanocrystalline magnetic materials such as Fe-Zr-B\cite{suzuki_high_1990}
and Fe-Si-B-M where M is an alloying addition\cite{yoshizawa_new_1988}. However, magnetic field annealing of medium- and high-entropy alloys, with the purpose of controlling atomic order, is currently an unexplored topic.

Perhaps the most widely studied high-entropy alloys are the family consisting of the Cantor alloy, CrMnFeCoNi, and its derivatives, collectively referred to as the Cantor-Wu alloys \cite{cantor_microstructural_2004, wu_recovery_2014, billington_bulk_2020}. 
Of particular interest is the three-component, medium-entropy CrCoNi, in which ASRO has been experimentally observed\cite{zhang_short-range_2020, li_evolution_2023}.
Also of interest is the widely-studied four-component CrFeCoNi alloy\cite{velisa_recent_2023}. 

A number of computational works based on first-principles calculations have examined ASRO in these materials. 
{ Tamm {\it et al.}\cite{tamm_atomic-scale_2015} studied the ternary CrCoNi alloy using lattice-based Monte Carlo (MC) simulations with energies evaluated via density functional theory (DFT).}
These calculations employed a spin-polarised scheme in which atoms are permitted to acquire magnetic moments and magnetically order, reflective of material behaviour at low temperatures.
They found that, at nearest neighbour distance,  Cr-Co and Co-Ni pairs were favoured, while Cr-Cr pairs were disfavoured. 
Ding {\it et al.}\cite{ding_tunable_2018} performed a subsequent study { on CrCoNi}, again using lattice-based MC simulations with energies evaluated via DFT.
These calculations were also spin polarised, and the results obtained by these authors for ASRO were in broad agreement with those of the earlier study.
Pei {\it et al.}\cite{pei_statistics_2020} described the internal energy of the CrCoNi system using an atomic cluster expansion, again giving results in agreement with earlier studies.
Further, these authors proposed a novel three-layered structure of lowest energy for the system, although whether this structure is truly the lowest energy configuration is disputed by Ghosh {\it et al.}\cite{ghosh_short-range_2022}, who instead find an MoPt$_2$-like ground state, using machine-learned (ML) interatomic potentials.
Another study employing ML interatomic potentials was performed by Du {\it et al.}\cite{du_chemical_2022}, highlighting competition between $\mathrm{L}1_2$-like and MoPt$_2$-like order in this material.
(The MoPt$_2$ and $\mathrm{L}1_2$ structures are visualised in Fig.~\ref{fig:partial_order}.)
The impact of magnetism on atomic ordering in this material has also been studied directly.
Ding {\it et al.}\cite{ding_tunable_2018} noted that the specifics of the ASRO depended on whether or not spin polarisation was invoked in their calculations, but no results for the unpolarised calculations are provided.
Walsh {\it et al.}\cite{walsh_magnetically_2021} also suggested that ASRO in CrCoNi is magnetically driven.
However, Ghosh {\it et al.}\cite{ghosh_short-range_2022} found that their MoPt$_2$-like ground state was not impacted by their treatment of magnetism in the system.

Tamm {\it et al.}\cite{tamm_atomic-scale_2015} also studied the quarternary CrFeCoNi alloy, finding similar atomic short-range order to that found in the CrCoNi system, with Cr-Co, Cr-Ni, Fe-Co, and Fe-Ni pairs favoured, and Cr-Cr and Fe-Fe and pairs disfavoured.
These results are in agreement with a study by Sch\"{o}nfeld {\it et al.}\cite{schonfeld_local_2019}, which used an approach based on the CPA to obtain effective pair interactions for atomistic modelling.
The impact of magnetism on atomic ordering in CrFeCoNi has also been noted.
Niu {\it et al.}\cite{niu_spin-driven_2015} suggested that magnetism was the driving force behind $\mathrm{L}1_2$-like ordering in this material, with Cr atoms preferring to couple antiferromagnetically with the other elements present and occupy the corners of the fcc lattice.
The hypothesis that atomic ordering is related to magnetism was also supported by another work by Fukushima {\it et al.}\cite{fukushima_local_2017}.

In our own earlier study\cite{woodgate_compositional_2022}, using an effective-medium-based (CPA), first-principles approach combined with a Landau-type linear response theory, we found results that are in good agreement with many of the aforementioned works, with the notable exception of the studies of CrFeCoNi conducted by Tamm {\it et al.} and Sch\"{o}nfeld {\it et al.}, where we instead found that Fe was only weakly correlated with the other elements present.
We qualitatively attributed this disagreement to the fact that our modelling described the alloy in a paramagnetic state, while the modelling techniques used in the earlier works allowed magnetic order to form.

In the present work, we use computational modelling to explicitly address the impact of the magnetic state on atomic short-range order in the aforementioned high- and medium-entropy alloys, by comparing the results of our linear response theory performed on the systems in their magnetically disordered state with results of the same analysis performed on systems in their magnetically ordered state.
We are able to show that, indeed, the competition between $\mathrm{L}1_2$-like and MoPt$_2$-like atomic order in CrCoNi is related to the material's magnetic state.
We also show that the magnetic state has a significant effect on the strength of correlations between Fe, Mn, and other $3d$ elements in the CrFeCoNi and CrMnFeCoNi systems.
These results have implications for materials modelling, by highlighting the importance of correct treatment of magnetism in these systems, as well as for materials processing; it may be possible to tune the nature and degree of ASRO in high-entropy alloy compositions by annealing samples in an applied magnetic field.

This paper is structured as follows. In section \ref{sec:theory}, we outline our methodology and the underlying theory.
Then, in section \ref{sec:results}, we give results of our calculations, and also give insight into the physical origins of predicted atomic order in terms of the materials' electronic structure.
Finally, in section \ref{sec:conculusions}, we summarise our results, give an outlook on their implications, and suggest possible further work.

\section{Theory}
\label{sec:theory}

\subsection{Linear Response Theory}
\label{sec:linear_response_theory}

Our technique for modelling compositional order in multicomponent alloys { centres around the two-point correlation function, an ASRO parameter. We use a Landau-type expansion of the free energy of the system to obtain this quantity {\it ab initio}}. Effects on the electronic structure and on the rearrangement of charge due to an applied inhomogeneous chemical perturbation are fully included. The inclusion of these effects is similar to the approach taken in Density Functional Perturbation Theory (DFPT)~\cite{savrasov_linear-response_1996, baroni_phonons_2001}, used to describe lattice dynamics {\it ab initio} and response functions for phonons, etc. Full details of our theory and extensive discussion can be found in earlier works\cite{khan_statistical_2016, woodgate_compositional_2022, woodgate_short-range_2023}. Our calculations assume a fixed ideal lattice, face-centred cubic (fcc) for the alloys studied in this paper, which represent the averaged atomic positions in the solid solution. The theory employed in this study is an extension of the $S^{(2)}$ theory developed for binary alloys \cite{gyorffy_concentration_1983, staunton_compositional_1994}. It has its groundings in statistical physics and in the seminal papers on concentration waves authored by Khachaturyan~\cite{khachaturyan_ordering_1978} and Gyorffy and Stocks~\cite{gyorffy_concentration_1983}.

A substitutional alloy with a fixed underlying lattice can be described by a set of site occupation numbers, $\{\xi_{i\alpha}\}$, where $\xi_{i\alpha}$=1 if site $i$ is occupied by an atom of species $\alpha$, and $\xi_{i\alpha}$=0 otherwise. The constraint that each lattice site be occupied by one atom and one atom only is expressed as $\sum_\alpha \xi_{i\alpha}$=1 for all lattice sites $i$. The overall concentration of each chemical species $\alpha$ is given by $c_\alpha = \frac{1}{N} \sum_i \xi_{i\alpha}$, where $N$ is the total number of lattice sites. It is natural to describe long-range order in such a system by the ensemble average of the site occupancies, writing $c_{i\alpha} = \langle \xi_{i\alpha} \rangle$, where we refer to $c_{i\alpha}$ as the site-wise concentrations. In the high-temperature, atomically disordered limit, these occupancies will be spatially homogeneous, taking the value of the overall concentration of that species. Below any atomic disorder-order transition, however, these occupancies will acquire a spatial dependence. The spatially dependent site-wise concentrations can be written as a fluctuation to the concentration distribution of the homogeneous system, $c_{i\alpha} = c_\alpha + \Delta c_{i\alpha}$. Given the underlying translational symmetry of the lattice, it is natural to write these fluctuations in reciprocal space using the so-called concentration wave formalism, pioneered by Khachaturayan\cite{khachaturyan_ordering_1978}. In this manner we write
\begin{equation}
c_{i\alpha} = c_\alpha + \sum_{\mathbf{k}} e^{i \mathbf{k} \cdot \mathbf{R}_i} \Delta c_\alpha(\mathbf{k})
\end{equation}
to describe a chemical fluctuation, where $\mathbf{R}_i$ is the lattice vector with corresponding occupancy $c_{i\alpha}$. As an example, we consider  $\mathrm{L}1_2$-type crystallographic order imposed on the fcc lattice, visualised in Fig.~\ref{fig:partial_order}. For an $A_3B$ binary system, $c_\alpha = (\frac{3}{4}, \frac{1}{4})$. The $\mathrm{L}1_2$-ordered structure, represented by atoms of species $B$ on the corners of the cubic unit cell, is then described by $\mathbf{k} = (0,0,1)$ and equivalent, with the change in concentration $\Delta c_\alpha = \frac{1}{\sqrt{2}} (-1, 1)$. It is clear that this formalism is extensible to multicomponent alloys.

Above any atomic ordering temperature, the natural quantity of interest is the two-point correlation function, {\it i.e.} the short-range order, written as
\begin{equation}
\Psi_{i\alpha j \alpha'} = \langle \xi_{i\alpha}\xi_{j\alpha'} \rangle - \langle \xi_{i\alpha} \rangle \langle \xi_{j\alpha'}\rangle,
\end{equation}
which assesses the degree to which different chemical species are spatially correlated in the material. This quantity is intrinsically related to the energetic cost of chemical fluctuations. Above any atomic ordering temperature, it is intuitive to think of chemical fluctuations positioned around the minimum of a high-dimensional bowl of a free energy surface. Information about how `steep' the various sides of this bowl are tells us which fluctuations are energetically costly, and which are energetically cheap. The ASRO in a material will be dominated by fluctuations which are energetically cheap.

To assess the energetic cost of a chemical fluctuation, we approximate the free energy, $\Omega$, of an alloy with inhomogeneous site-wise concentration distribution, $\{c_{i\alpha}\}$, by
\begin{equation}
    \Omega^{(1)}[\{\nu_{i\alpha}\}, \{c_{i\alpha}\}] = -\frac{1}{\beta} \sum_{i\alpha} c_{i\alpha} \ln c_{i\alpha}
    - \sum_{i\alpha} \nu_{i\alpha} c_{i\alpha} + \langle \Omega_\text{el} \rangle_0 [\{c_{i\alpha}\}],
\end{equation}
where the three terms on the right-hand side of Eq. (2) read left to right describe entropic contributions, site-wise chemical potentials, and an average of the electronic contribution to the free energy of the system, respectively. We then make a Landau-type expansion of the free energy of the system around a homogeneous reference state, {\it i.e} the atomically disordered solid solution, writing
\begin{align}
    \Omega^{(1)}(\{c_{i\alpha}\}) =& \; \Omega^{(1)}(\{c_{\alpha}\}) + \sum_{i\alpha} \frac{\partial \Omega^{(1)}}{\partial c_{i\alpha}} \Big\vert_{\{c_{\alpha}\}} \Delta c_{i\alpha} \nonumber \\ 
    &+ \frac{1}{2} \sum_{i\alpha; j\alpha'} \frac{\partial^2 \Omega^{(1)}}{\partial c_{i\alpha} \partial c_{j\alpha'}} \Big\vert_{\{c_{\alpha}\}} \Delta c_{i\alpha}\Delta c_{j\alpha'} + \dots.
\label{eq:landau}   
\end{align} 
The site-wise chemical potentials serve as Lagrange multipliers in the linear response theory, but as their variation is not considered to be relevant to the underlying physics, terms involving these derivatives are dropped \cite{khan_statistical_2016, woodgate_compositional_2022, woodgate_short-range_2023}. The symmetry of the high-temperature, homogeneous state - the solid solution - and the requirement that any imposed fluctuation conserves the overall concentrations of each chemical species ensures that the first-order term vanishes. To second order, the change in free energy, $\delta \Omega^{(1)}$ as a result of a fluctuation is therefore written
\begin{equation}
    \delta \Omega^{(1)} = \frac{1}{2} \sum_{i\alpha; j\alpha'} \Delta c_{i\alpha} [\beta^{-1} \, C_{\alpha\alpha'}^{-1} - S^{(2)}_{i\alpha, j\alpha'}] \Delta c_{j\alpha'},
    \label{eq:chemical_stability_real}
\end{equation}
where $C_{\alpha \alpha'}^{-1} = \frac{\delta_{\alpha \alpha'}}{c_\alpha}$ is associated with the entropic contributions, and the term $-\frac{\partial^2 \langle \Omega_\text{el} \rangle_0}{\partial c_{i\alpha} \partial c_{j\alpha'}} \equiv S^{(2)}_{i\alpha;j\alpha'}$ is the second-order concentration derivative of the average energy of the disordered alloy. The evaluation of this term has been covered in depth in earlier works \cite{khan_statistical_2016, woodgate_compositional_2022} and we omit discussion of it here for brevity.

Crucially, $S^{(2)}_{i\alpha;j\alpha'}$, is evaluated in reciprocal space in our codes, and therefore the change in free energy of Eq.~\ref{eq:chemical_stability_real} is Fourier-transformed as:
\begin{equation}
    \delta \Omega^{(1)} = \frac{1}{2} \sum_{\bf k} \sum_{\alpha, \alpha'} \Delta c_\alpha({\bf k}) [\beta^{-1} C^{-1}_{\alpha \alpha'} -S^{(2)}_{\alpha \alpha'}({\bf k})] \Delta c_{\alpha'}({\bf k}).
\label{eq:chemical_stability_reciprocal}
\end{equation}
The matrix in square brackets $[\beta^{-1} C^{-1}_{\alpha \alpha'} -S^{(2)}_{\alpha \alpha'}({\bf k})]$, referred to as the chemical stability matrix, is related to an estimate of the ASRO, $\Psi_{i\alpha;j\alpha'}$. When searching for an disorder-order transition, we consider decreasing temperature and look for the temperature at which the lowest lying eigenvalue of this matrix passes through zero for any $\mathbf{k}$-vector in the irreducible Brillouin Zone. When this eigenvalue passes through zero at some temperature $T_\text{us}$ and wavevector $\mathbf{k}_\text{us}$, we infer the presence of an disorder-order transition with chemical polarisation $\Delta c_\alpha$ given by the associated eigenvector. In this fashion we can predict both dominant ASRO and also the temperature at which the solid solution becomes unstable and a (partially) ordered phase emerges.

\subsection{Atomistic Modelling}
\label{sec:atomistic}

To further explore the phase space of a given multicomponent alloy system, it is possible to map the concentration derivatives of the internal energy of the alloy, $S^{(2)}_{\alpha\alpha'}(\mathbf{k})$, to a pairwise real-space interaction. This real-space model is lattice-based and has the conventional Bragg-Williams Hamiltonian\cite{bragg_effect_1934, bragg_effect_1935},
\begin{equation}
    H = \frac{1}{2}\sum_{i \alpha; j\alpha'} V_{i\alpha; j\alpha'} \xi_{i \alpha} \xi_{j \alpha'} + \sum_{i\alpha} \nu_{\alpha} \xi_{i\alpha}.
    \label{eq:b-w}
\end{equation}
The effective pairwise interactions, $V_{i\alpha; j\alpha'}$, are recovered from $S^{(2)}_{\alpha\alpha'}(\mathbf{k})$ by means of a backwards Fourier transform, with the mapping from reciprocal-space to real-space and fixing of the gauge degree of freedom as specified in earlier works\cite{khan_statistical_2016, woodgate_compositional_2022, woodgate_short-range_2023}.

{We note that, because the current {\it ab initio} framework evaluates a pair correlation function, a pair interaction is obtained. In principle, however, the methodology is extensible to evaluation of higher-order correlations and the corresponding many-body interactions, akin to a cluster expansion. We suggest that one approach which could be followed to obtain these would be analogous to that used in Refs.~\cite{mendive-tapia_theory_2017} and \cite{mendive_tapia_ab_2020}, which obtained higher-order interactions in a magnetic setting within a similar first-principles framework to that of the present work.

We also emphasise that this model assumes a fixed, underlying lattice, and does not directly account for local lattice distortions. However, for the family of alloys considered in this work, which form on the fcc lattice, local lattice distortions are known to be small\cite{oh_lattice_2016}. We therefore believe that this lattice-based model is appropriate.}

To investigate the phase behaviour of these systems with this {pairwise} atomistic model, we use the Metropolis Monte-Carlo algorithm with Kawasaki dynamics\cite{landau_guide_2014}. In this approach the chemical potential term, $\sum_{i\alpha} \nu_{\alpha} \xi_{i\alpha}$, in Eq.~\ref{eq:b-w} is dropped, as these dynamics naturally conserve overall concentrations of each chemical species by permitting only swaps of pairs of atoms.

The algorithm proceeds as follows: the occupation numbers are randomly initialised with the overall number of atoms of each species, establishing the concentrations, as the only restriction. A pair of atomic sites is selected at random, and the change in internal energy $\Delta H$ realized from swapping the site occupancies is calculated. If the change in energy is negative ($\Delta H < 0$) the swap is accepted unconditionally, while if the change is positive ($\Delta H > 0$) the swap is accepted with acceptance probability $e^{-\beta \Delta H}$. To assess the configurational contribution to the specific heat capacity (SHC) of the system, we use the fluctuation-dissipation theorem\cite{allen_computer_2017}. At thermodynamic equilibrium, this theorem allows us to estimate the specific heat capacity as
\begin{equation}
    C = \frac{1}{k_b T^2} \left( \langle E^2 \rangle - \langle E \rangle^2 \right),
\end{equation}
to obtain our SHC curves.

To quantify ASRO in our simulations, we generate the Warren-Cowley ASRO parameters\cite{cowley_approximate_1950, cowley_short-range_1965}, $\alpha^{pq}_n$, adapted to the multicomponent alloy setting, 
\begin{equation}
    \alpha^{pq}_n = 1 - \frac{P^{pq}_n}{c_q},
\end{equation}
where $n$ refers to the $n$th coordination shell, $P^{pq}_n$ is the conditional probability of an atom of type $q$ neighbouring an atom of type $p$ on coordination shell $n$, and $c_q$ is the overall concentration of atom type $q$. When $\alpha^{pq}_n>0$, $p$-$q$ pairs are disfavoured on shell $n$, while when $\alpha^{pq}_n<0$ they are favoured. The value 0 corresponds to the ideal, random, solid solution.

\section{Results and Discussion}
\label{sec:results}

\subsection{Electronic Structure Calculations}
\label{sec:electronic_structure}

To model the electronic `glue' bonding atoms together and driving ASRO, we first generate the self-consistent, single-electron potentials of density functional theory (DFT) \cite{martin_electronic_2004}, which are used as the basis for performing linear response calculations. These potentials are generated in the Korringa-Kohn-Rostoker (KKR) formulation of DFT, using the coherent potential approximation (CPA) to produce an effective medium reflecting the average electronic structure of the high-symmetry, disordered solid solution \cite{faulkner_calculating_1980, faulkner_multiple_2018, johnson_total-energy_1990}. We use the all-electron HUTSEPOT code \cite{hoffmann_magnetic_2020} to generate these potentials, although a number of other KKR-CPA codes would also be suitable. We perform spin-polarised, scalar-relativistic calculations within the atomic sphere approximation (ASA)\cite{stocks_complete_1978}, employing an angular momentum cutoff of $l_\text{max} = 3$ for basis set expansions, a $20\times20\times20$ Monkhorst-Pack grid\cite{monkhorst_special_1976} for integrals over the Brillouin zone, and a 24 point semi-circular Gauss-Legendre grid in the complex plane to integrate over valence energies. We use the local density approximation (LDA) and the exchange-correlation functional is that of Perdew-Wang\cite{perdew_accurate_1992}. Cubic lattice parameters of 3.56, 3.57 and 3.59\AA \, are used for CrCoNi, CrFeCoNi, and CrMnFeCoNi respectively, consistent with their experimental values \cite{zhang_influence_2015, yin_yield_2020, cantor_microstructural_2004}. The magnetically disordered (paramagnetic) states are described within the Disordered Local Moment (DLM) picture\cite{gyorffy_first-principles_1985}.

\begin{figure*}[p]
\centering
\includegraphics[width=0.9\textwidth]{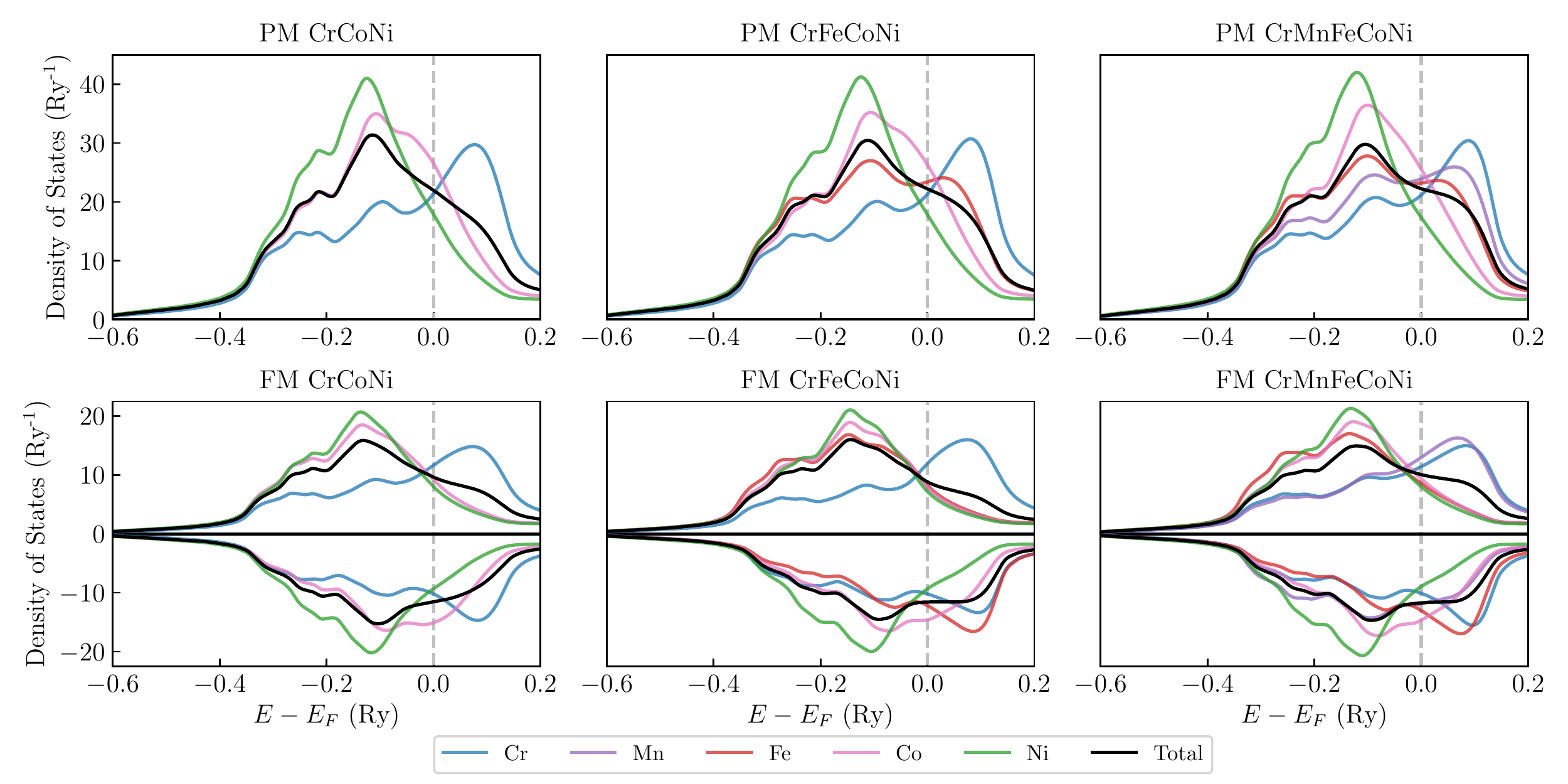}
\caption{Plots of the total and species-resolved density of states for the three systems, in both magnetically disordered (`PM') and magnetically ordered (`FM') states. The total DoS is given by the weighted average of the species-resolved curves. It can be seen that there are significant shifts to the DoS for elements acquiring a large magnetic moment in the magnetically ordered calculations. In the majority spin channel of CrCoNi, for instance, Co starts to `look' like Ni, while in CrMnFeCoNi, Mn makes a similar contribution to Cr to the total DoS.}
\label{fig:dos_comparison}
\end{figure*}

Figure \ref{fig:dos_comparison} shows a comparison of the total density of states (DoS) for the disordered solid solution for the three alloys considered in this work, with the top row providing the total DoS in the magnetically disordered state within the DLM picture, and the bottom row providing the spin-polarised DoS for the magnetically ordered state. As in our earlier work, we emphasise that we believe it is most physically correct to model these systems as magnetically disordered (paramagnetic). All three studied alloys have Curie temperatures that are below room temperature\cite{billington_bulk_2020}, and well below any experimental annealing temperature. In this current work, we highlight the impact of magnetism on predicted ASRO by performing the same calculations as were conducted in our earlier work\cite{woodgate_compositional_2022}, but this time employing potentials describing a magnetically ordered state.

In the magnetically disordered (paramagnetic) state, we find that only Mn and Fe support local moments within the DLM picture, while Cr, Co, and Ni do not. For the potentials that model a magnetically ordered state, the resultant magnitudes of the magnetic moments for each chemical species are tabulated in Table \ref{table:mag_moments}. It is noted that the magnetically ordered states are ferrimagnetic; our calculations indicate that moments on Fe, Co, and Ni align with the total magnetic moment of the system, while those on Cr and Mn are found to anti-align, in good agreement with earlier DFT results and experimental data\cite{billington_bulk_2020, walsh_magnetically_2021, ghosh_short-range_2022}. Also consistent with the experimental data are the computed small energy differences between paramagnetic and magnetically ordered states, indicative of low Curie (or N\'eel) temperatures. In all cases we find that the magnetically ordered state is of lower energy than that of the paramagnetic state, with energy differences per lattice site of 0.192 mRy, 1.047 mRy, and 0.310 mRy for CrCoNi, CrFeCoNi, and CrMnFeCoNi respectively.

Compared to the results obtained for the paramagnetic potentials, which we discussed in our earlier work\cite{woodgate_compositional_2022}, we find that allowing the system to establish magnetic order significantly alters the intuitive picture of how various elements in the systems might interact. This conclusion is illustrated by considering the three-component system, CrCoNi. The large positive moment on Co is shown by the shift to the left (towards lower energy) of its DoS curve in the majority spin channel to lie almost on top of the Ni curve. In this channel, the electronic structure begins to resemble that of CrNi$_2$, a compound that is known to adopt the MoPt$_2$ structure\cite{rahaman_first-principles_2014}. For the four-component system, the species-resolved DoS contribution from Fe is also significantly altered, suggesting that the nature of interactions involving Fe will change when the chemical stability analysis is performed. Finally, for CrMnFeCoNi, the negative moment on Mn shifts its DoS curve in the majority spin channel to lie almost on top of that of Cr, while in the minority channel it lies almost on top of the averaged total DoS curve.

\begin{table}[p]
\begin{ruledtabular}
\centering
\begin{tabular}{l|lllll}
\multirow{2}{*}{Material} & \multicolumn{5}{l}{Elemental Moment ($\mu_B$)}        \\ \cline{2-6} 
                          & Cr       & Mn       & Fe    & Co       & Ni       \\ \hline
CrCoNi               & $-$0.163 &          &       &  0.832   & 0.148 \\
CrFeCoNi             & $-$0.651 &          & 1.922 & 1.079    & 0.282 \\
CrMnFeCoNi           & $-$0.154 & $-$1.177 & 1.786 & 0.773    & 0.133
\end{tabular}
\end{ruledtabular}
\caption{Average magnetic moments associated with each chemical species in the magnetically ordered solid solution. Notably, in CrMnFeCoNi, Mn prefers to anti-align with the other elements present.}
\label{table:mag_moments}
\end{table}

\subsection{Linear Response Analysis}
\label{sec:linear_response}

\begin{figure*}[p]
\centering
\includegraphics[width=0.9\textwidth]{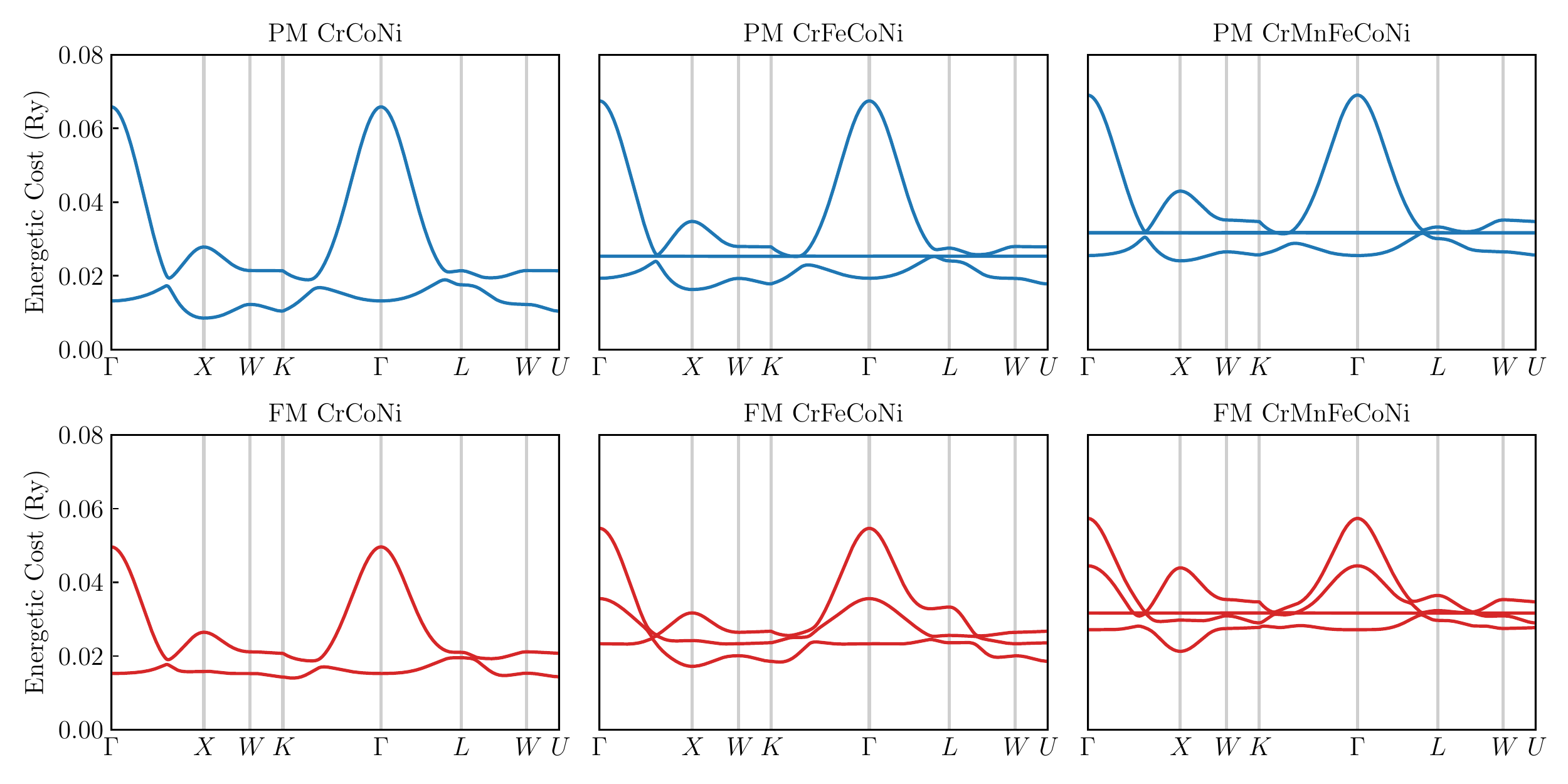}
\caption{Plots of the eigenvalues of the chemical stability matrix around the irreducible Brillouin Zone for  the three systems, generated from both the magnetically disordered (`PM') and magnetically ordered (`FM') potentials. It can be seen that the nature of modes in all three materials are altered when the magnetic symmetry is broken; there is more mixing between modes and previously flat modes acquire more distinct shape, indicative of strengthening correlations. Most notably, in CrCoNi, there is clear competition between minima, i.e. the dominant SRO. From magnetically disordered to ordered states, the location of the minimum shifts from $\mathbf{k} = (0,0,1)$, {\it i.e.} the $X$ point, to $\mathbf{k} = (0, \frac{2}{3}, \frac{2}{3})$, a point along the line from $\Gamma$ to $K$. This is indicative of competition between $\mathrm{L}1_2$-like and MoPt$_2$-like order.}
\label{fig:eigenvalue_comparison}
\end{figure*}

We perform linear response calculations as described in Sec.~\ref{sec:linear_response_theory} for all three alloys using the self-consistent potentials describing both magnetically ordered (ferrimagnetic) and magnetically disordered (paramagnetic) states. Plots of the eigenvalues of the chemical stability matrices are shown in Fig.~\ref{fig:eigenvalue_comparison}. In our earlier work that employed only paramagnetic calculations, we portrayed a simple picture of Cr-Co interactions as dominant, while Mn and Fe interact only weakly, contributing near-flat concentration wave modes and diluting the strength of interactions\cite{woodgate_compositional_2022}. (A near-flat concentration wave mode is associated with weak interactions and, therefore, with weak ASRO.) The picture for the systems modelled assuming a magnetically ordered state differs significantly from this earlier work; the flat concentration wave mode associated with Fe in CrFeCoNi is completely changed in character, and in CrMnFeCoNi there is only one flat mode now, not two. This lifting of degeneracy of the two flat modes is associated with stronger atomic correlations between Fe, Mn and other elements present.

In Fig.~\ref{fig:eigenvalue_comparison}, for the ternary CrCoNi alloy, clear competition can be seen between energetic minima at  $\mathbf{k} = (0,0,1)$, indicative of $\mathrm{L}1_2$-like order, and $\mathbf{k} = (0, \frac{2}{3}, \frac{2}{3})$, associated with MoPt$_2$-like order. When modelling the magnetically ordered state, the minimum at $(0, \frac{2}{3}, \frac{2}{3})$ sits lower than that at $(0,0,1)$, but for the magnetically disordered state it is the minimum at $(0,0,1)$ which is at lower energy and therefore describes the dominant ASRO. This outcome highlights the competition between the two different types of order in this system, as has been noted in earlier works\cite{du_chemical_2022, ghosh_short-range_2022}, but also emphasises the connection between this competition and magnetism, as discussed by Walsh {\it et al.}\cite{walsh_magnetically_2021}.

\begin{figure*}[p]
\centering
\begin{subfigure}{0.49\textwidth}
\centering
\includegraphics[height=0.6\textwidth]{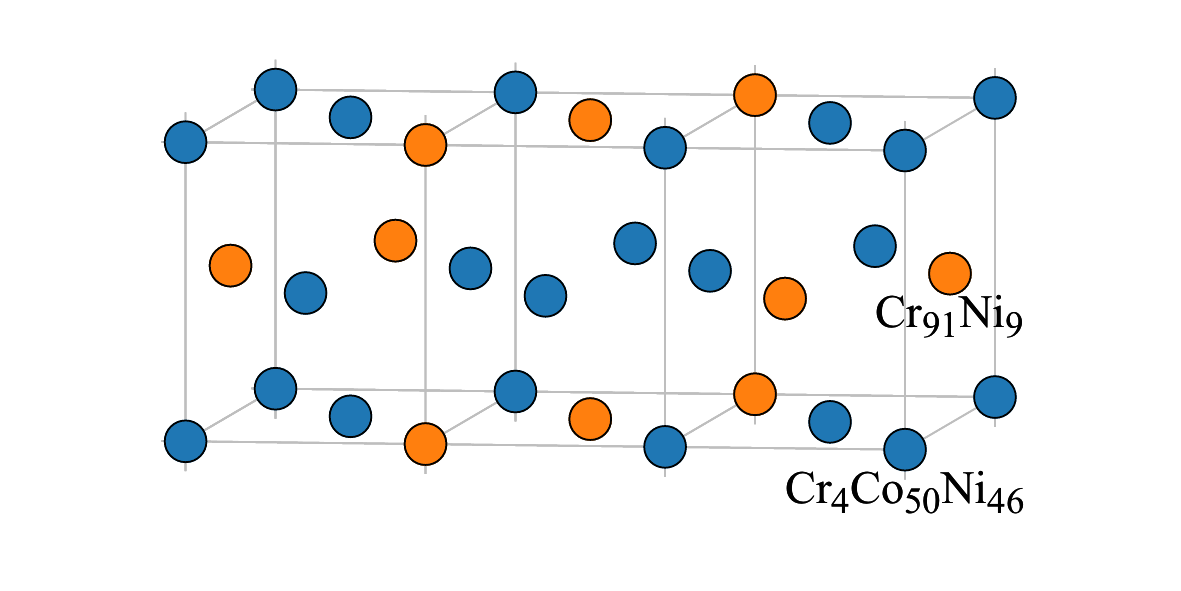}
\caption{FM CrCoNi---MoPt$_2$ Structure}
\end{subfigure}
\begin{subfigure}{0.49\textwidth}
\centering
\includegraphics[height=0.6\textwidth]{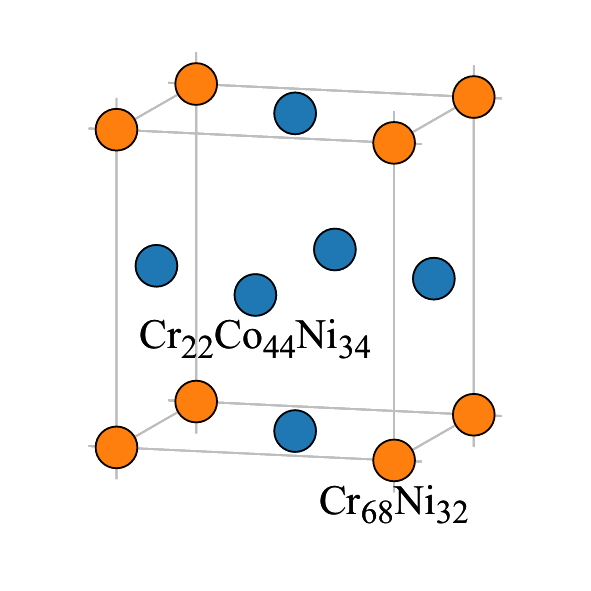}
\caption{PM CrCoNi---$\mathrm{L}1_2$ Structure}
\end{subfigure}
\caption{Comparison of the partially ordered structures predicted with our linear response theory for CrCoNi in both its magnetically ordered (`FM') and magnetically disordered (`PM') states. Both orderings are driven by Cr, but in the magnetically ordered state, MoPt$_2$-like order is predicted, while in the paramagnetic state it is $\mathrm{L}1_2$-like order that is favoured.}
\label{fig:partial_order}
\end{figure*}

Predicted compositional ordering temperatures, concentration wave modes, and chemical polarisations for the alloys, assuming both magnetically ordered and disordered states, are tabulated in Table \ref{table:linear_response}, and the corresponding partially ordered structures for CrCoNi are illustrated in Figure \ref{fig:partial_order}. Structures with partial atomic ordering are obtained by allowing the concentration wave predicted by the chemical instability analysis to `grow' until one sublattice contains (at least) one atomic species whose concentration reaches zero. This condition identifies the the largest permitted chemical fluctuation consistent with that polarisation. 

The most striking change in predicted atomic ordering is noted for the ternary CrCoNi alloy,  where the calculated atomic disorder-order temperature for a magnetically ordered state is significantly lower than that for the same system with magnetic disorder. In this case the predicted structure is now MoPt$_2$-like rather than $\mathrm{L}1_2$-like. In the four-component CrFeCoNi alloy, while the wave-vector describing the order remains the same, magnetic order is found to change the chemical polarisation of the mode significantly, strengthening correlations involving Fe atoms and weakening those involving Co atoms. These results are now in good agreement with the study by Tamm {\it et al.}\cite{tamm_atomic-scale_2015}, where Monte Carlo simulations with energies evaluated via spin-polarised DFT calculations found strong correlations between Fe and other elements, a result which we failed to find when modelling the paramagnetic state. A similar story holds true in the complex CrMnFeCoNi alloy, where correlations involving both Fe and Mn, which both acquire large magnetic moments, are also strengthened compared to the results obtained in our earlier work modelling the paramagnetic state\cite{woodgate_compositional_2022}.

One notable disagreement between our results and those of earlier works is described in the recent study by  Ghosh {\it et al.}\cite{ghosh_short-range_2022}, which predicted MoPt$_2$-like order emerging at 975 K in CrCoNi. Simulations using low-rank interatomic potentials have previously been in good agreement with results obtained using our approach when modelling refractory HEAs\cite{kostiuchenko_impact_2019, woodgate_short-range_2023}, so this discrepancy is surprising. We note that a recent experimental study by Li {\it et al.} \cite{li_evolution_2023} performed annealing of CrCoNi at a variety of temperatures, including one sample annealed at 673 K for 500 h. They found evidence of increasing ASRO in their samples, but no evidence of a long-range ordered state was reported. Further, the known MoPt$_2$-former, CrNi$_2$, has an experimentally observed atomic ordering temperature of 863 K, which is in good agreement with the theoretical prediction around 880 K \cite{staunton_compositional_1994, rahaman_first-principles_2014}. We would expect the atomic ordering temperature for the ternary CrCoNi alloy to be lower than that predicted for CrNi$_2$, because the increased entropy of mixing of the ternary alloy over that of the binary alloy will stabilise the solid solution.

\begin{table*}[p]
\scriptsize
\begin{ruledtabular}
\centering
\begin{tabular}{ll|rcrrrrr}
Material & Magnetic State& $T_\text{us} (K) $ & $\mathbf{k}_\text{us}$ & $\Delta c_1$ & $\Delta c_2$ & $\Delta c_3$ & $\Delta c_4$ & $\Delta c_5$ \\ \hline
CrCoNi & Paramagnetic     &    606 & $(0, 0, 1)$       & $0.724$  & $-0.689$ & $-0.035$ &&                \\
 & Ordered   &    252 & $(0, \frac{2}{3}, \frac{2}{3})$       &  $0.813$ & $-0.468$ & $-0.345$ &&                \\ \hline
CrFeCoNi & Paramagnetic  &    404 & $(0, 0, 1)$       & $0.723$ & $-0.048$ & $-0.689$ & $0.014$ &         \\ 
& Ordered &    313 & $(0, 0, 1)$       &  $0.531$ & $-0.787$ & $-0.052$ & $0.308$ &         \\ \hline
CrMnFeCoNi & Paramagnetic &     281 & $(0, 0, 1)$ &  $0.723$ & $0.011$ & $-0.082$ & $-0.685$ & $0.033$        \\
 & Ordered &   319 & $(0, 0, 1)$ &  $0.227$ & $-0.625$ & $0.706$ & $-0.077$ & $-0.231$        \\
\end{tabular}
\end{ruledtabular}
\caption{Transition temperatures, modes, and chemical polarisations for the considered systems generated from both the paramagnetic (DLM) potentials, and magnetically ordered potentials. The values for the paramagnetic potentials are taken from Ref.~\citenum{woodgate_compositional_2022}. Elements are numbered as per the specified composition, e.g. for CrFeCoNi Cr=1, Fe=2, Co=3, and Ni=4. There are differences in the temperature and nature of predicted compositional order between the two magnetic states. Notably, for CrCoNi, the mode shifts from $\mathbf{k} = (0,0,1)$ to $\mathbf{k} = (0, \frac{2}{3}, \frac{2}{3})$, a point along the line from $\Gamma$ to $K$, indicative of a shift from $\mathrm{L}1_2$-like to MoPt$_2$-like order}\label{table:linear_response}%
\end{table*}

\subsection{Atomistic Modelling}

To understand further the nature of atomic short-range order present below any initial ordering temperature in these complex HEA materials, we fit a real-space interaction to our reciprocal space $S^{(2)}_{\alpha\alpha'}(\mathbf{k})$ data (representing internal energy derivatives) to recover a model suitable for lattice-based atomistic modelling, as outlined in \ref{sec:atomistic}. { We sample $S^{(2)}_{\alpha\alpha'}(\mathbf{k})$ at 56 $\mathbf{k}$-points around the irreducible Brillouin zone, including high-symmetry points. We find that fitting $V_{i\alpha; j\alpha'}$ to the first four coordination shells of the fcc lattice (consistent with Ref.~\cite{woodgate_compositional_2022}) captures the $S^{(2)}_{\alpha\alpha'}(\mathbf{k})$ data with good accuracy. We assume that interactions are isotropic and homogeneous, and write $V^{(n)}_{\alpha \alpha'}$ to denote the interaction between species $\alpha$ and $\alpha'$ on coordination shell $n$.} Our {fitted} pairwise atom-atom interchange parameters, $V^{(n)}_{\alpha \alpha'}$, are provided in the supplemental material\cite{supplementary}. 

Lattice-based Monte Carlo simulations were performed using a system consisting of $8\times8\times8$ cubic unit cells, each with 4 lattice sites per unit cell, for a total of 2048 atoms. Periodic boundary conditions were applied. The systems were prepared in an initially random configuration, then computationally annealed from 1200 K to 10 K in steps of 1 K, with $10^4$ MC steps per atom at each temperature.

\begin{figure*}[p]
\centering
\includegraphics[width=\textwidth]{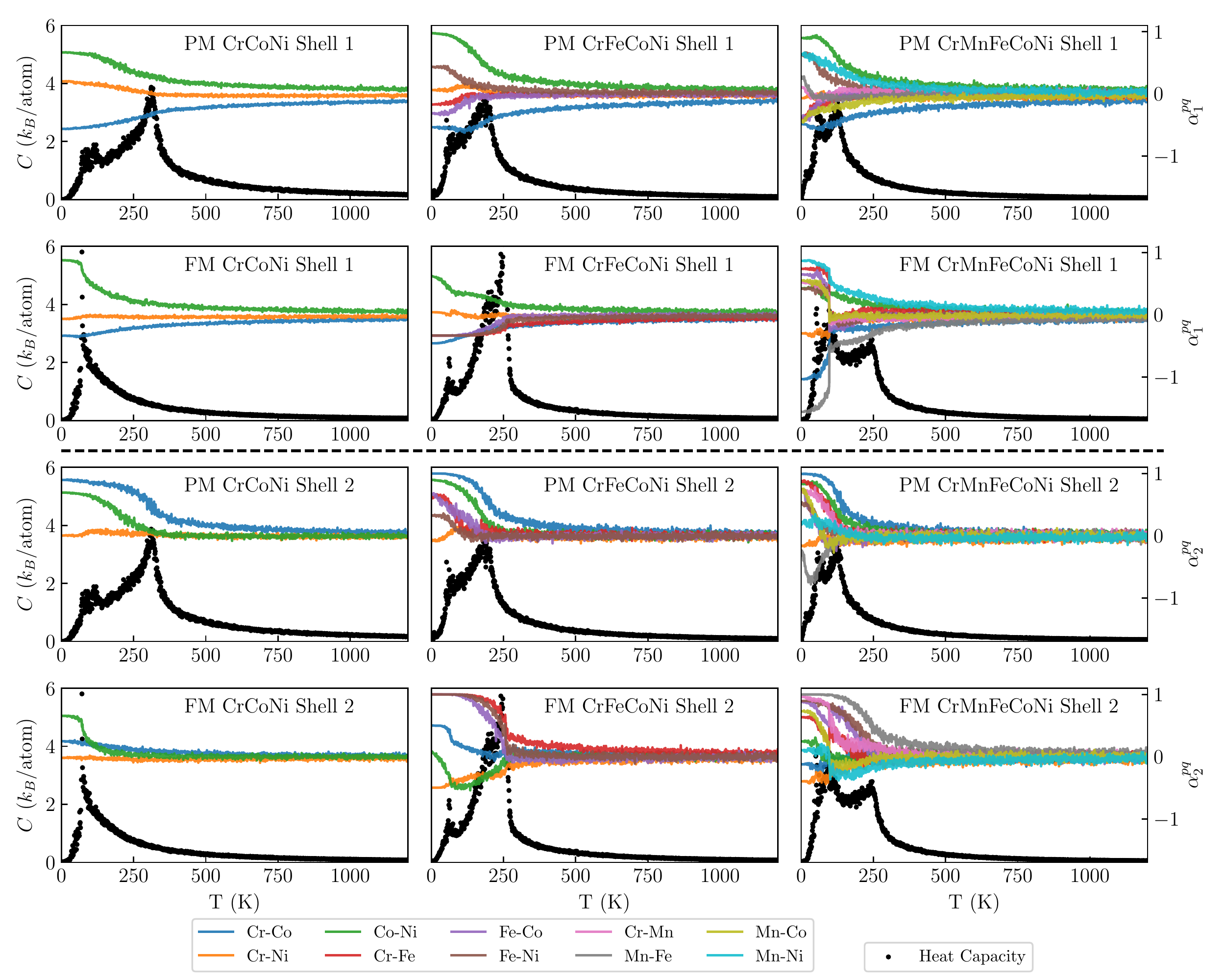}
\caption{Plots of the Warren-Cowley short-range order parameters on first and second coordination shells for the three considered systems, along with a measure of the system's specific heat capacity, {$C$}. Each sub-plot is labelled to indicate the alloy and magnetic state for which it shows results. The magnetically ordered results show a significant weakening of ordering between Cr and Co, and a strengthening of interactions between Fe and the other elements.}
\label{fig:monte_carlo}
\end{figure*}

Figure~\ref{fig:monte_carlo} compares the Warren-Cowley ASRO parameters on first and second coordination shells, {\it i.e.} for first- and second-nearest neighbours. Results are shown for interactions computed based on the magnetically disodered (paramagnetic) state, as well as for those computed based on the magnetically ordered (ferrimagnetic) state. There are clear differences in the strength and nature of predicted ASRO in all three alloy compositions, depending on which magnetic state is simulated. Although still favored, the Cr-Co correlations in CrCoNi are significantly weakened when the system is modeled under the assumption of magnetic order. The favouring of Cr-Co pairs on the first neighbour shell is consistent with the results of earlier computational studies\cite{tamm_atomic-scale_2015, ding_tunable_2018, ghosh_short-range_2022, du_chemical_2022, walsh_magnetically_2021}. In CrFeCoNi, Cr-Co correlations are also weakened, while correlations between Fe and all three other elements are significantly strengthened. This outcome now agrees qualitatively with the work by Tamm {\it et al.}\cite{tamm_atomic-scale_2015} as well as with other previous works \cite{schonfeld_local_2019, niu_spin-driven_2015, fukushima_local_2017}. Finally, while the picture is less clear-cut for CrMnFeCoNi, it can be seen that there are significant qualitative differences in ASRO between magnetically ordered and magnetically disordered states, with Mn-Fe and Mn-Ni atomic correlations dominant in the low temperature regime when the system is modelled in a magnetically ordered state. We note that there is little information in the literature to allow assessment of our results obtained for the five-component alloy system, as the huge potential configuration space challenges most conventional supercell-based computational techniques.

These results have clear implications for the materials modelling community as well as for experimental researchers. It is confirmed that due consideration must be is given to the magnetic state of a material when modelling multicomponent alloy systems containing $3d$ transition metals. The presence and nature of magnetic order significantly impacts the strength and nature of predicted ASRO. It may also be true that there are some high-entropy alloy compositions for which the application of a magnetic field during the annealing process, especially at temperatures below any magnetic transition temperature(s), will impact the ASRO by affecting the material's magnetic state and in turn impact its mechanical properties.

\section{Conclusions}
\label{sec:conculusions}

In summary, results have been presented that shed light on the importance of correct treatment of a system's magnetic state when modelling multicomponent alloys containing $3d$ transition metals. Depending on how the magnetic state is computationally treated, significant differences in predictions of short- and long-range order are noted for all three multicomponent alloy systems considered in this study. In particular, our results reiterate that competition between $\mathrm{L}1_2$-like and MoPt$_2$-like order in CrCoNi is unequivocally connected to magnetism in the system, as has been suggested in earlier works.

Our described approach for performing computationally inexpensive modelling of magnetic alloys in both magnetically ordered and magnetically disordered states has potential to inform experiment by suggesting whether multicomponent alloys with sufficiently high Curie temperatures should be annealed above or below $T_\text{C}$, or in an applied magnetic field, to promote desired ordering and/or other beneficial materials properties. We are in the process of studying some such systems to better understand the connections between magnetism and ASRO. We also note that our Monte Carlo simulations performed using this simple, pairwise Hamiltonian can quickly provide atomic configurations with physically motivated ASRO for use in supercell calculations and in training datasets for machine-learned interatomic potentials.

Our results therefore have implications both for materials modelling as well as for materials preparation. An interesting avenue of further study on the CrCoNi system would be to model how magnetic interactions in this system vary in different atomically ordered structures when compared to the disordered solid solution to further elucidate the connection between magnetism and ASRO. Another avenue of work to pursue experimentally would be to anneal some medium- and high-entropy alloy compositions in an applied magnetic field to assess the degree to which ASRO, and therefore materials properties, can be tuned.

\begin{acknowledgments}
We gratefully acknowledge the support of the UK Engineering and Physical Sciences Research Council, Grant No. EP/W021331/1. C.D.W. is supported by a studentship within the UK Engineering and Physical Sciences Research Council-supported Centre for Doctoral Training in Modelling of Heterogeneous Systems, Grant No. EP/S022848/1. This work was also supported in part by the U.S. Department of Energy, Office of Basic Energy Sciences under Award Number DE SC0022168 and by the U.S. National Science Foundation under Award ID 2118164.
\end{acknowledgments}


\begin{thebibliography}{69}%
\makeatletter
\providecommand \@ifxundefined [1]{%
 \@ifx{#1\undefined}
}%
\providecommand \@ifnum [1]{%
 \ifnum #1\expandafter \@firstoftwo
 \else \expandafter \@secondoftwo
 \fi
}%
\providecommand \@ifx [1]{%
 \ifx #1\expandafter \@firstoftwo
 \else \expandafter \@secondoftwo
 \fi
}%
\providecommand \natexlab [1]{#1}%
\providecommand \enquote  [1]{``#1''}%
\providecommand \bibnamefont  [1]{#1}%
\providecommand \bibfnamefont [1]{#1}%
\providecommand \citenamefont [1]{#1}%
\providecommand \href@noop [0]{\@secondoftwo}%
\providecommand \href [0]{\begingroup \@sanitize@url \@href}%
\providecommand \@href[1]{\@@startlink{#1}\@@href}%
\providecommand \@@href[1]{\endgroup#1\@@endlink}%
\providecommand \@sanitize@url [0]{\catcode `\\12\catcode `\$12\catcode
  `\&12\catcode `\#12\catcode `\^12\catcode `\_12\catcode `\%12\relax}%
\providecommand \@@startlink[1]{}%
\providecommand \@@endlink[0]{}%
\providecommand \url  [0]{\begingroup\@sanitize@url \@url }%
\providecommand \@url [1]{\endgroup\@href {#1}{\urlprefix }}%
\providecommand \urlprefix  [0]{URL }%
\providecommand \Eprint [0]{\href }%
\providecommand \doibase [0]{https://doi.org/}%
\providecommand \selectlanguage [0]{\@gobble}%
\providecommand \bibinfo  [0]{\@secondoftwo}%
\providecommand \bibfield  [0]{\@secondoftwo}%
\providecommand \translation [1]{[#1]}%
\providecommand \BibitemOpen [0]{}%
\providecommand \bibitemStop [0]{}%
\providecommand \bibitemNoStop [0]{.\EOS\space}%
\providecommand \EOS [0]{\spacefactor3000\relax}%
\providecommand \BibitemShut  [1]{\csname bibitem#1\endcsname}%
\let\auto@bib@innerbib\@empty
\bibitem [{\citenamefont {Zhang}\ \emph {et~al.}(2014)\citenamefont {Zhang},
  \citenamefont {Zuo}, \citenamefont {Tang}, \citenamefont {Gao}, \citenamefont
  {Dahmen}, \citenamefont {Liaw},\ and\ \citenamefont
  {Lu}}]{zhang_microstructures_2014}%
  \BibitemOpen
  \bibfield  {author} {\bibinfo {author} {\bibfnamefont {Y.}~\bibnamefont
  {Zhang}}, \bibinfo {author} {\bibfnamefont {T.~T.}\ \bibnamefont {Zuo}},
  \bibinfo {author} {\bibfnamefont {Z.}~\bibnamefont {Tang}}, \bibinfo {author}
  {\bibfnamefont {M.~C.}\ \bibnamefont {Gao}}, \bibinfo {author} {\bibfnamefont
  {K.~A.}\ \bibnamefont {Dahmen}}, \bibinfo {author} {\bibfnamefont {P.~K.}\
  \bibnamefont {Liaw}},\ and\ \bibinfo {author} {\bibfnamefont {Z.~P.}\
  \bibnamefont {Lu}},\ }\href {https://doi.org/10.1016/j.pmatsci.2013.10.001}
  {\bibfield  {journal} {\bibinfo  {journal} {Progress in Materials Science}\
  }\textbf {\bibinfo {volume} {61}},\ \bibinfo {pages} {1} (\bibinfo {year}
  {2014})}\BibitemShut {NoStop}%
\bibitem [{\citenamefont {Pickering}\ and\ \citenamefont
  {Jones}(2016)}]{pickering_high-entropy_2016}%
  \BibitemOpen
  \bibfield  {author} {\bibinfo {author} {\bibfnamefont {E.~J.}\ \bibnamefont
  {Pickering}}\ and\ \bibinfo {author} {\bibfnamefont {N.~G.}\ \bibnamefont
  {Jones}},\ }\href {https://doi.org/10.1080/09506608.2016.1180020} {\bibfield
  {journal} {\bibinfo  {journal} {International Materials Reviews}\ }\textbf
  {\bibinfo {volume} {61}},\ \bibinfo {pages} {183} (\bibinfo {year} {2016})}\BibitemShut {NoStop}%
\bibitem [{\citenamefont {Gao}\ \emph {et~al.}(2016)\citenamefont {Gao},
  \citenamefont {Yeh}, \citenamefont {Liaw},\ and\ \citenamefont
  {Zhang}}]{gao_high-entropy_2016}%
  \BibitemOpen
  \bibinfo {editor} {\bibfnamefont {M.~C.}\ \bibnamefont {Gao}}, \bibinfo
  {editor} {\bibfnamefont {J.-W.}\ \bibnamefont {Yeh}}, \bibinfo {editor}
  {\bibfnamefont {P.~K.}\ \bibnamefont {Liaw}},\ and\ \bibinfo {editor}
  {\bibfnamefont {Y.}~\bibnamefont {Zhang}},\ eds.,\ \href
  {https://doi.org/10.1007/978-3-319-27013-5} {\emph {\bibinfo {title}
  {High-{Entropy} {Alloys}}}}\ (\bibinfo  {publisher} {Springer International
  Publishing},\ \bibinfo {address} {Cham},\ \bibinfo {year} {2016})\BibitemShut
  {NoStop}%
\bibitem [{\citenamefont {Miracle}\ and\ \citenamefont
  {Senkov}(2017)}]{miracle_critical_2017}%
  \BibitemOpen
  \bibfield  {author} {\bibinfo {author} {\bibfnamefont {D.~B.}\ \bibnamefont
  {Miracle}}\ and\ \bibinfo {author} {\bibfnamefont {O.~N.}\ \bibnamefont
  {Senkov}},\ }\href {https://doi.org/10.1016/j.actamat.2016.08.081} {\bibfield
   {journal} {\bibinfo  {journal} {Acta Materialia}\ }\textbf {\bibinfo
  {volume} {122}},\ \bibinfo {pages} {448} (\bibinfo {year}
  {2017})}\BibitemShut {NoStop}%
\bibitem [{\citenamefont {George}\ \emph {et~al.}(2019)\citenamefont {George},
  \citenamefont {Raabe},\ and\ \citenamefont
  {Ritchie}}]{george_high-entropy_2019}%
  \BibitemOpen
  \bibfield  {author} {\bibinfo {author} {\bibfnamefont {E.~P.}\ \bibnamefont
  {George}}, \bibinfo {author} {\bibfnamefont {D.}~\bibnamefont {Raabe}},\ and\
  \bibinfo {author} {\bibfnamefont {R.~O.}\ \bibnamefont {Ritchie}},\ }\href
  {https://doi.org/10.1038/s41578-019-0121-4} {\bibfield  {journal} {\bibinfo
  {journal} {Nature Reviews Materials}\ }\textbf {\bibinfo {volume} {4}},\
  \bibinfo {pages} {515} (\bibinfo {year} {2019})}\BibitemShut {NoStop}%
\bibitem [{\citenamefont {Singh}\ \emph {et~al.}(2015)\citenamefont {Singh},
  \citenamefont {Smirnov},\ and\ \citenamefont {Johnson}}]{singh_atomic_2015}%
  \BibitemOpen
  \bibfield  {author} {\bibinfo {author} {\bibfnamefont {P.}~\bibnamefont
  {Singh}}, \bibinfo {author} {\bibfnamefont {A.~V.}\ \bibnamefont {Smirnov}},\
  and\ \bibinfo {author} {\bibfnamefont {D.~D.}\ \bibnamefont {Johnson}},\
  }\href {https://doi.org/10.1103/PhysRevB.91.224204} {\bibfield  {journal}
  {\bibinfo  {journal} {Physical Review B}\ }\textbf {\bibinfo {volume} {91}},\
  \bibinfo {pages} {224204} (\bibinfo {year} {2015})}\BibitemShut {NoStop}%
\bibitem [{\citenamefont {Troparevsky}\ \emph {et~al.}(2015)\citenamefont
  {Troparevsky}, \citenamefont {Morris}, \citenamefont {Kent}, \citenamefont
  {Lupini},\ and\ \citenamefont {Stocks}}]{troparevsky_criteria_2015}%
  \BibitemOpen
  \bibfield  {author} {\bibinfo {author} {\bibfnamefont {M.~C.}\ \bibnamefont
  {Troparevsky}}, \bibinfo {author} {\bibfnamefont {J.~R.}\ \bibnamefont
  {Morris}}, \bibinfo {author} {\bibfnamefont {P.~R.~C.}\ \bibnamefont {Kent}},
  \bibinfo {author} {\bibfnamefont {A.~R.}\ \bibnamefont {Lupini}},\ and\
  \bibinfo {author} {\bibfnamefont {G.~M.}\ \bibnamefont {Stocks}},\ }\href
  {https://doi.org/10.1103/PhysRevX.5.011041} {\bibfield  {journal} {\bibinfo
  {journal} {Phys. Rev. X}\ }\textbf {\bibinfo {volume} {5}},\ \bibinfo {pages}
  {011041} (\bibinfo {year} {2015})}\BibitemShut {NoStop}%
\bibitem [{\citenamefont {Liu}\ \emph {et~al.}(2016)\citenamefont {Liu},
  \citenamefont {Lindwall}, \citenamefont {Gheno},\ and\ \citenamefont
  {Liu}}]{liu_thermodynamic_2016}%
  \BibitemOpen
  \bibfield  {author} {\bibinfo {author} {\bibfnamefont {X.~L.}\ \bibnamefont
  {Liu}}, \bibinfo {author} {\bibfnamefont {G.}~\bibnamefont {Lindwall}},
  \bibinfo {author} {\bibfnamefont {T.}~\bibnamefont {Gheno}},\ and\ \bibinfo
  {author} {\bibfnamefont {Z.-K.}\ \bibnamefont {Liu}},\ }\href
  {https://doi.org/10.1016/j.calphad.2015.12.007} {\bibfield  {journal}
  {\bibinfo  {journal} {Calphad}\ }\textbf {\bibinfo {volume} {52}},\ \bibinfo
  {pages} {125} (\bibinfo {year} {2016})}\BibitemShut {NoStop}%
\bibitem [{\citenamefont {Gao}\ \emph {et~al.}(2017)\citenamefont {Gao},
  \citenamefont {Zhang}, \citenamefont {Gao}, \citenamefont {Zhang},
  \citenamefont {Ouyang}, \citenamefont {Widom},\ and\ \citenamefont
  {Hawk}}]{gao_thermodynamics_2017}%
  \BibitemOpen
  \bibfield  {author} {\bibinfo {author} {\bibfnamefont {M.}~\bibnamefont
  {Gao}}, \bibinfo {author} {\bibfnamefont {C.}~\bibnamefont {Zhang}}, \bibinfo
  {author} {\bibfnamefont {P.}~\bibnamefont {Gao}}, \bibinfo {author}
  {\bibfnamefont {F.}~\bibnamefont {Zhang}}, \bibinfo {author} {\bibfnamefont
  {L.}~\bibnamefont {Ouyang}}, \bibinfo {author} {\bibfnamefont
  {M.}~\bibnamefont {Widom}},\ and\ \bibinfo {author} {\bibfnamefont
  {J.}~\bibnamefont {Hawk}},\ }\href
  {https://doi.org/10.1016/j.cossms.2017.08.001} {\bibfield  {journal}
  {\bibinfo  {journal} {Current Opinion in Solid State and Materials Science}\
  }\textbf {\bibinfo {volume} {21}},\ \bibinfo {pages} {238} (\bibinfo {year}
  {2017})}\BibitemShut {NoStop}%
\bibitem [{\citenamefont {Fernández-Caballero}\ \emph
  {et~al.}(2017)\citenamefont {Fernández-Caballero}, \citenamefont {Wróbel},
  \citenamefont {Mummery},\ and\ \citenamefont
  {Nguyen-Manh}}]{fernandez-caballero_short-range_2017}%
  \BibitemOpen
  \bibfield  {author} {\bibinfo {author} {\bibfnamefont {A.}~\bibnamefont
  {Fernández-Caballero}}, \bibinfo {author} {\bibfnamefont {J.~S.}\
  \bibnamefont {Wróbel}}, \bibinfo {author} {\bibfnamefont {P.~M.}\
  \bibnamefont {Mummery}},\ and\ \bibinfo {author} {\bibfnamefont
  {D.}~\bibnamefont {Nguyen-Manh}},\ }\href
  {https://doi.org/10.1007/s11669-017-0582-3} {\bibfield  {journal} {\bibinfo
  {journal} {Journal of Phase Equilibria and Diffusion}\ }\textbf {\bibinfo
  {volume} {38}},\ \bibinfo {pages} {391} (\bibinfo {year} {2017})}\BibitemShut
  {NoStop}%
\bibitem [{\citenamefont {Widom}(2018)}]{widom_modeling_2018}%
  \BibitemOpen
  \bibfield  {author} {\bibinfo {author} {\bibfnamefont {M.}~\bibnamefont
  {Widom}},\ }\href {https://doi.org/10.1557/jmr.2018.222} {\bibfield
  {journal} {\bibinfo  {journal} {Journal of Materials Research}\ }\textbf
  {\bibinfo {volume} {33}},\ \bibinfo {pages} {2881} (\bibinfo {year}
  {2018})}\BibitemShut {NoStop}%
\bibitem [{\citenamefont {Ikeda}\ \emph {et~al.}(2019)\citenamefont {Ikeda},
  \citenamefont {Grabowski},\ and\ \citenamefont {Körmann}}]{ikeda_ab_2019}%
  \BibitemOpen
  \bibfield  {author} {\bibinfo {author} {\bibfnamefont {Y.}~\bibnamefont
  {Ikeda}}, \bibinfo {author} {\bibfnamefont {B.}~\bibnamefont {Grabowski}},\
  and\ \bibinfo {author} {\bibfnamefont {F.}~\bibnamefont {Körmann}},\ }\href
  {https://doi.org/10.1016/j.matchar.2018.06.019} {\bibfield  {journal}
  {\bibinfo  {journal} {Materials Characterization}\ }\textbf {\bibinfo
  {volume} {147}},\ \bibinfo {pages} {464} (\bibinfo {year}
  {2019})}\BibitemShut {NoStop}%
\bibitem [{\citenamefont {Kostiuchenko}\ \emph {et~al.}(2019)\citenamefont
  {Kostiuchenko}, \citenamefont {Körmann}, \citenamefont {Neugebauer},\ and\
  \citenamefont {Shapeev}}]{kostiuchenko_impact_2019}%
  \BibitemOpen
  \bibfield  {author} {\bibinfo {author} {\bibfnamefont {T.}~\bibnamefont
  {Kostiuchenko}}, \bibinfo {author} {\bibfnamefont {F.}~\bibnamefont
  {Körmann}}, \bibinfo {author} {\bibfnamefont {J.}~\bibnamefont
  {Neugebauer}},\ and\ \bibinfo {author} {\bibfnamefont {A.}~\bibnamefont
  {Shapeev}},\ }\href {https://doi.org/10.1038/s41524-019-0195-y} {\bibfield
  {journal} {\bibinfo  {journal} {npj Computational Materials}\ }\textbf
  {\bibinfo {volume} {5}},\ \bibinfo {pages} {55} (\bibinfo {year}
  {2019})}\BibitemShut {NoStop}%
\bibitem [{\citenamefont {Ferrari}\ \emph {et~al.}(2020)\citenamefont
  {Ferrari}, \citenamefont {Dutta}, \citenamefont {Gubaev}, \citenamefont
  {Ikeda}, \citenamefont {Srinivasan}, \citenamefont {Grabowski},\ and\
  \citenamefont {Körmann}}]{ferrari_frontiers_2020}%
  \BibitemOpen
  \bibfield  {author} {\bibinfo {author} {\bibfnamefont {A.}~\bibnamefont
  {Ferrari}}, \bibinfo {author} {\bibfnamefont {B.}~\bibnamefont {Dutta}},
  \bibinfo {author} {\bibfnamefont {K.}~\bibnamefont {Gubaev}}, \bibinfo
  {author} {\bibfnamefont {Y.}~\bibnamefont {Ikeda}}, \bibinfo {author}
  {\bibfnamefont {P.}~\bibnamefont {Srinivasan}}, \bibinfo {author}
  {\bibfnamefont {B.}~\bibnamefont {Grabowski}},\ and\ \bibinfo {author}
  {\bibfnamefont {F.}~\bibnamefont {Körmann}},\ }\href
  {https://doi.org/10.1063/5.0025310} {\bibfield  {journal} {\bibinfo
  {journal} {Journal of Applied Physics}\ }\textbf {\bibinfo {volume} {128}},\
  \bibinfo {pages} {150901} (\bibinfo {year} {2020})}\BibitemShut {NoStop}%
\bibitem [{\citenamefont {Wang}\ \emph {et~al.}(2021)\citenamefont {Wang},
  \citenamefont {Hu}, \citenamefont {Huang},\ and\ \citenamefont
  {Zheng}}]{wang_thermodynamic_2021}%
  \BibitemOpen
  \bibfield  {author} {\bibinfo {author} {\bibfnamefont {P.}~\bibnamefont
  {Wang}}, \bibinfo {author} {\bibfnamefont {B.}~\bibnamefont {Hu}}, \bibinfo
  {author} {\bibfnamefont {X.}~\bibnamefont {Huang}},\ and\ \bibinfo {author}
  {\bibfnamefont {C.}~\bibnamefont {Zheng}},\ }\href
  {https://doi.org/10.1016/j.calphad.2021.102252} {\bibfield  {journal}
  {\bibinfo  {journal} {Calphad}\ }\textbf {\bibinfo {volume} {73}},\ \bibinfo
  {pages} {102252} (\bibinfo {year} {2021})}\BibitemShut {NoStop}%
\bibitem [{\citenamefont {Yeh}\ \emph {et~al.}(2004)\citenamefont {Yeh},
  \citenamefont {Chen}, \citenamefont {Lin}, \citenamefont {Gan}, \citenamefont
  {Chin}, \citenamefont {Shun}, \citenamefont {Tsau},\ and\ \citenamefont
  {Chang}}]{yeh_nanostructured_2004}%
  \BibitemOpen
  \bibfield  {author} {\bibinfo {author} {\bibfnamefont {J.-W.}\ \bibnamefont
  {Yeh}}, \bibinfo {author} {\bibfnamefont {S.-K.}\ \bibnamefont {Chen}},
  \bibinfo {author} {\bibfnamefont {S.-J.}\ \bibnamefont {Lin}}, \bibinfo
  {author} {\bibfnamefont {J.-Y.}\ \bibnamefont {Gan}}, \bibinfo {author}
  {\bibfnamefont {T.-S.}\ \bibnamefont {Chin}}, \bibinfo {author}
  {\bibfnamefont {T.-T.}\ \bibnamefont {Shun}}, \bibinfo {author}
  {\bibfnamefont {C.-H.}\ \bibnamefont {Tsau}},\ and\ \bibinfo {author}
  {\bibfnamefont {S.-Y.}\ \bibnamefont {Chang}},\ }\href
  {https://doi.org/10.1002/adem.200300567} {\bibfield  {journal} {\bibinfo
  {journal} {Advanced Engineering Materials}\ }\textbf {\bibinfo {volume}
  {6}},\ \bibinfo {pages} {299} (\bibinfo {year} {2004})}\BibitemShut {NoStop}%
\bibitem [{\citenamefont {Zhang}\ \emph {et~al.}(2017)\citenamefont {Zhang},
  \citenamefont {Zhao}, \citenamefont {Jin}, \citenamefont {Xue}, \citenamefont
  {Velisa}, \citenamefont {Bei}, \citenamefont {Huang}, \citenamefont {Ko},
  \citenamefont {Pagan}, \citenamefont {Neuefeind}, \citenamefont {Weber},\
  and\ \citenamefont {Zhang}}]{zhang_local_2017}%
  \BibitemOpen
  \bibfield  {author} {\bibinfo {author} {\bibfnamefont {F.~X.}~\bibnamefont
  {Zhang}}, \bibinfo {author} {\bibfnamefont {S.}~\bibnamefont {Zhao}},
  \bibinfo {author} {\bibfnamefont {K.}~\bibnamefont {Jin}}, \bibinfo {author}
  {\bibfnamefont {H.}~\bibnamefont {Xue}}, \bibinfo {author} {\bibfnamefont
  {G.}~\bibnamefont {Velisa}}, \bibinfo {author} {\bibfnamefont
  {H.}~\bibnamefont {Bei}}, \bibinfo {author} {\bibfnamefont {R.}~\bibnamefont
  {Huang}}, \bibinfo {author} {\bibfnamefont {J.~Y.~P.}~\bibnamefont {Ko}}, \bibinfo
  {author} {\bibfnamefont {D.~C.}~\bibnamefont {Pagan}}, \bibinfo {author}
  {\bibfnamefont {J.~C.}~\bibnamefont {Neuefeind}}, \bibinfo {author}
  {\bibfnamefont {W.~J.}~\bibnamefont {Weber}},\ and\ \bibinfo {author}
  {\bibfnamefont {Y.}~\bibnamefont {Zhang}},\ }\href
  {https://doi.org/10.1103/PhysRevLett.118.205501} {\bibfield  {journal}
  {\bibinfo  {journal} {Physical Review Letters}\ }\textbf {\bibinfo {volume}
  {118}},\ \bibinfo {pages} {205501} (\bibinfo {year} {2017})}\BibitemShut
  {NoStop}%
\bibitem [{\citenamefont {Zhang}\ \emph {et~al.}(2020)\citenamefont {Zhang},
  \citenamefont {Zhao}, \citenamefont {Ding}, \citenamefont {Chong},
  \citenamefont {Jia}, \citenamefont {Ophus}, \citenamefont {Asta},
  \citenamefont {Ritchie},\ and\ \citenamefont
  {Minor}}]{zhang_short-range_2020}%
  \BibitemOpen
  \bibfield  {author} {\bibinfo {author} {\bibfnamefont {R.}~\bibnamefont
  {Zhang}}, \bibinfo {author} {\bibfnamefont {S.}~\bibnamefont {Zhao}},
  \bibinfo {author} {\bibfnamefont {J.}~\bibnamefont {Ding}}, \bibinfo {author}
  {\bibfnamefont {Y.}~\bibnamefont {Chong}}, \bibinfo {author} {\bibfnamefont
  {T.}~\bibnamefont {Jia}}, \bibinfo {author} {\bibfnamefont {C.}~\bibnamefont
  {Ophus}}, \bibinfo {author} {\bibfnamefont {M.}~\bibnamefont {Asta}},
  \bibinfo {author} {\bibfnamefont {R.~O.}\ \bibnamefont {Ritchie}},\ and\
  \bibinfo {author} {\bibfnamefont {A.~M.}\ \bibnamefont {Minor}},\ }\href
  {https://doi.org/10.1038/s41586-020-2275-z} {\bibfield  {journal} {\bibinfo
  {journal} {Nature}\ }\textbf {\bibinfo {volume} {581}},\ \bibinfo {pages}
  {283} (\bibinfo {year} {2020})}\BibitemShut {NoStop}%
\bibitem [{\citenamefont {Li}\ \emph {et~al.}(2023)\citenamefont {Li},
  \citenamefont {Chen}, \citenamefont {Kuroiwa}, \citenamefont {Ito},
  \citenamefont {Yuge}, \citenamefont {Kishida}, \citenamefont {Tanimoto},
  \citenamefont {Yu}, \citenamefont {Inui},\ and\ \citenamefont
  {George}}]{li_evolution_2023}%
  \BibitemOpen
  \bibfield  {author} {\bibinfo {author} {\bibfnamefont {L.}~\bibnamefont
  {Li}}, \bibinfo {author} {\bibfnamefont {Z.}~\bibnamefont {Chen}}, \bibinfo
  {author} {\bibfnamefont {S.}~\bibnamefont {Kuroiwa}}, \bibinfo {author}
  {\bibfnamefont {M.}~\bibnamefont {Ito}}, \bibinfo {author} {\bibfnamefont
  {K.}~\bibnamefont {Yuge}}, \bibinfo {author} {\bibfnamefont {K.}~\bibnamefont
  {Kishida}}, \bibinfo {author} {\bibfnamefont {H.}~\bibnamefont {Tanimoto}},
  \bibinfo {author} {\bibfnamefont {Y.}~\bibnamefont {Yu}}, \bibinfo {author}
  {\bibfnamefont {H.}~\bibnamefont {Inui}},\ and\ \bibinfo {author}
  {\bibfnamefont {E.~P.}\ \bibnamefont {George}},\ }\href
  {https://doi.org/10.1016/j.actamat.2022.118537} {\bibfield  {journal}
  {\bibinfo  {journal} {Acta Materialia}\ }\textbf {\bibinfo {volume} {243}},\
  \bibinfo {pages} {118537} (\bibinfo {year} {2023})}\BibitemShut {NoStop}%
\bibitem [{\citenamefont {Wu}\ \emph {et~al.}(2021)\citenamefont {Wu},
  \citenamefont {Zhang}, \citenamefont {Yuan}, \citenamefont {Huang},
  \citenamefont {Wen}, \citenamefont {Wang}, \citenamefont {Zhang},
  \citenamefont {Wu}, \citenamefont {Liu}, \citenamefont {Wang}, \citenamefont
  {Jiang},\ and\ \citenamefont {Lu}}]{wu_short-range_2021}%
  \BibitemOpen
  \bibfield  {author} {\bibinfo {author} {\bibfnamefont {Y.}~\bibnamefont
  {Wu}}, \bibinfo {author} {\bibfnamefont {F.}~\bibnamefont {Zhang}}, \bibinfo
  {author} {\bibfnamefont {X.}~\bibnamefont {Yuan}}, \bibinfo {author}
  {\bibfnamefont {H.}~\bibnamefont {Huang}}, \bibinfo {author} {\bibfnamefont
  {X.}~\bibnamefont {Wen}}, \bibinfo {author} {\bibfnamefont {Y.}~\bibnamefont
  {Wang}}, \bibinfo {author} {\bibfnamefont {M.}~\bibnamefont {Zhang}},
  \bibinfo {author} {\bibfnamefont {H.}~\bibnamefont {Wu}}, \bibinfo {author}
  {\bibfnamefont {X.}~\bibnamefont {Liu}}, \bibinfo {author} {\bibfnamefont
  {H.}~\bibnamefont {Wang}}, \bibinfo {author} {\bibfnamefont {S.}~\bibnamefont
  {Jiang}},\ and\ \bibinfo {author} {\bibfnamefont {Z.}~\bibnamefont {Lu}},\
  }\href {https://doi.org/10.1016/j.jmst.2020.06.018} {\bibfield  {journal}
  {\bibinfo  {journal} {Journal of Materials Science \& Technology}\ }\textbf
  {\bibinfo {volume} {62}},\ \bibinfo {pages} {214} (\bibinfo {year}
  {2021})}\BibitemShut {NoStop}%
\bibitem [{\citenamefont {Ziehl}\ \emph {et~al.}(2023)\citenamefont {Ziehl},
  \citenamefont {Morris},\ and\ \citenamefont {Zhang}}]{ziehl_detection_2023}%
  \BibitemOpen
  \bibfield  {author} {\bibinfo {author} {\bibfnamefont {T.~J.}\ \bibnamefont
  {Ziehl}}, \bibinfo {author} {\bibfnamefont {D.}~\bibnamefont {Morris}},\ and\
  \bibinfo {author} {\bibfnamefont {P.}~\bibnamefont {Zhang}},\ }\href
  {https://doi.org/10.1016/j.isci.2023.106209} {\bibfield  {journal} {\bibinfo
  {journal} {iScience}\ ,\ \bibinfo {pages} {106209}} (\bibinfo {year}
  {2023})}\BibitemShut {NoStop}%
\bibitem [{\citenamefont {Guthe}(1897)}]{guthe_influence_1897}%
  \BibitemOpen
  \bibfield  {author} {\bibinfo {author} {\bibfnamefont {K.~E.}\ \bibnamefont
  {Guthe}},\ }\href {https://doi.org/10.1109/T-AIEE.1897.5570176} {\bibfield
  {journal} {\bibinfo  {journal} {Transactions of the American Institute of
  Electrical Engineers}\ }\textbf {\bibinfo {volume} {XIV}},\ \bibinfo {pages}
  {55} (\bibinfo {year} {1897})}\BibitemShut {NoStop}%
\bibitem [{\citenamefont {Pender}\ and\ \citenamefont
  {Jones}(1913)}]{pender_annealing_1913}%
  \BibitemOpen
  \bibfield  {author} {\bibinfo {author} {\bibfnamefont {H.}~\bibnamefont
  {Pender}}\ and\ \bibinfo {author} {\bibfnamefont {R.~L.}\ \bibnamefont
  {Jones}},\ }\href {https://doi.org/10.1103/PhysRev.1.259} {\bibfield
  {journal} {\bibinfo  {journal} {Physical Review}\ }\textbf {\bibinfo {volume}
  {1}},\ \bibinfo {pages} {259} (\bibinfo {year} {1913})}\BibitemShut {NoStop}%
\bibitem [{\citenamefont {Fuji}(2009)}]{fuji_hiromichi_effects_2009}%
  \BibitemOpen
  \bibfield  {author} {\bibinfo {author} {\bibfnamefont {H.}~\bibnamefont
  {Fuji}},\ }\emph {\bibinfo {title} {Effects of {Magnetic}-{Field} on
  {Elemental} {Process} for {Microstructural} {Development} of {Iron}-{Based}
  {Polycrystalline} {Materials}}},\ \href
  {https://www.phase-trans.msm.cam.ac.uk/2011/Doctoral_thesis_fujii.pdf} {Ph.D.
  thesis},\ \bibinfo  {school} {Tohuku University}, \bibinfo {address} {Japan}
  (\bibinfo {year} {2009})\BibitemShut {NoStop}%
\bibitem [{\citenamefont {McCurrie}(1982)}]{mccurrie_chapter_1982}%
  \BibitemOpen
  \bibfield  {author} {\bibinfo {author} {\bibfnamefont {R.}~\bibnamefont
  {McCurrie}},\ }in\ \href {https://doi.org/10.1016/S1574-9304(05)80089-6}
  {\emph {\bibinfo {booktitle} {Handbook of {Ferromagnetic} {Materials}}}},\
  Vol.~\bibinfo {volume} {3}\ (\bibinfo  {publisher} {Elsevier},\ \bibinfo
  {year} {1982})\ pp.\ \bibinfo {pages} {107--188}\BibitemShut {NoStop}%
\bibitem [{\citenamefont {Suzuki}\ \emph {et~al.}(1990)\citenamefont {Suzuki},
  \citenamefont {Kataoka}, \citenamefont {Inoue}, \citenamefont {Makino},\ and\
  \citenamefont {Masumoto}}]{suzuki_high_1990}%
  \BibitemOpen
  \bibfield  {author} {\bibinfo {author} {\bibfnamefont {K.}~\bibnamefont
  {Suzuki}}, \bibinfo {author} {\bibfnamefont {N.}~\bibnamefont {Kataoka}},
  \bibinfo {author} {\bibfnamefont {A.}~\bibnamefont {Inoue}}, \bibinfo
  {author} {\bibfnamefont {A.}~\bibnamefont {Makino}},\ and\ \bibinfo {author}
  {\bibfnamefont {T.}~\bibnamefont {Masumoto}},\ }\href
  {https://doi.org/10.2320/matertrans1989.31.743} {\bibfield  {journal}
  {\bibinfo  {journal} {Materials Transactions, JIM}\ }\textbf {\bibinfo
  {volume} {31}},\ \bibinfo {pages} {743} (\bibinfo {year} {1990})}\BibitemShut
  {NoStop}%
\bibitem [{\citenamefont {Yoshizawa}\ \emph {et~al.}(1988)\citenamefont
  {Yoshizawa}, \citenamefont {Oguma},\ and\ \citenamefont
  {Yamauchi}}]{yoshizawa_new_1988}%
  \BibitemOpen
  \bibfield  {author} {\bibinfo {author} {\bibfnamefont {Y.}~\bibnamefont
  {Yoshizawa}}, \bibinfo {author} {\bibfnamefont {S.}~\bibnamefont {Oguma}},\
  and\ \bibinfo {author} {\bibfnamefont {K.}~\bibnamefont {Yamauchi}},\ }\href
  {https://doi.org/10.1063/1.342149} {\bibfield  {journal} {\bibinfo  {journal}
  {Journal of Applied Physics}\ }\textbf {\bibinfo {volume} {64}},\ \bibinfo
  {pages} {6044} (\bibinfo {year} {1988})}\BibitemShut {NoStop}%
\bibitem [{\citenamefont {Cantor}\ \emph {et~al.}(2004)\citenamefont {Cantor},
  \citenamefont {Chang}, \citenamefont {Knight},\ and\ \citenamefont
  {Vincent}}]{cantor_microstructural_2004}%
  \BibitemOpen
  \bibfield  {author} {\bibinfo {author} {\bibfnamefont {B.}~\bibnamefont
  {Cantor}}, \bibinfo {author} {\bibfnamefont {I.~T.~H.}\ \bibnamefont
  {Chang}}, \bibinfo {author} {\bibfnamefont {P.}~\bibnamefont {Knight}},\ and\
  \bibinfo {author} {\bibfnamefont {A.~J.~B.}\ \bibnamefont {Vincent}},\ }\href
  {https://doi.org/10.1016/j.msea.2003.10.257} {\bibfield  {journal} {\bibinfo
  {journal} {Materials Science and Engineering: A}\ }\textbf {\bibinfo {volume}
  {375-377}},\ \bibinfo {pages} {213} (\bibinfo {year} {2004})}\BibitemShut
  {NoStop}%
\bibitem [{\citenamefont {Wu}\ \emph {et~al.}(2014)\citenamefont {Wu},
  \citenamefont {Bei}, \citenamefont {Otto}, \citenamefont {Pharr},\ and\
  \citenamefont {George}}]{wu_recovery_2014}%
  \BibitemOpen
  \bibfield  {author} {\bibinfo {author} {\bibfnamefont {Z.}~\bibnamefont
  {Wu}}, \bibinfo {author} {\bibfnamefont {H.}~\bibnamefont {Bei}}, \bibinfo
  {author} {\bibfnamefont {F.}~\bibnamefont {Otto}}, \bibinfo {author}
  {\bibfnamefont {G.~M.}\ \bibnamefont {Pharr}},\ and\ \bibinfo {author}
  {\bibfnamefont {E.~P.}\ \bibnamefont {George}},\ }\href
  {https://doi.org/10.1016/j.intermet.2013.10.024} {\bibfield  {journal}
  {\bibinfo  {journal} {Intermetallics}\ }\textbf {\bibinfo {volume} {46}},\
  \bibinfo {pages} {131} (\bibinfo {year} {2014})}\BibitemShut {NoStop}%
\bibitem [{\citenamefont {Billington}\ \emph {et~al.}(2020)\citenamefont
  {Billington}, \citenamefont {James}, \citenamefont {Harris-Lee},
  \citenamefont {Lagos}, \citenamefont {O'Neill}, \citenamefont {Tsuda},
  \citenamefont {Toyoki}, \citenamefont {Kotani}, \citenamefont {Nakamura},
  \citenamefont {Bei}, \citenamefont {Mu}, \citenamefont {Samolyuk},
  \citenamefont {Stocks}, \citenamefont {Duffy}, \citenamefont {Taylor},
  \citenamefont {Giblin},\ and\ \citenamefont
  {Dugdale}}]{billington_bulk_2020}%
  \BibitemOpen
  \bibfield  {author} {\bibinfo {author} {\bibfnamefont {D.}~\bibnamefont
  {Billington}}, \bibinfo {author} {\bibfnamefont {A.~D.~N.}\ \bibnamefont
  {James}}, \bibinfo {author} {\bibfnamefont {E.~I.}\ \bibnamefont
  {Harris-Lee}}, \bibinfo {author} {\bibfnamefont {D.~A.}\ \bibnamefont
  {Lagos}}, \bibinfo {author} {\bibfnamefont {D.}~\bibnamefont {O'Neill}},
  \bibinfo {author} {\bibfnamefont {N.}~\bibnamefont {Tsuda}}, \bibinfo
  {author} {\bibfnamefont {K.}~\bibnamefont {Toyoki}}, \bibinfo {author}
  {\bibfnamefont {Y.}~\bibnamefont {Kotani}}, \bibinfo {author} {\bibfnamefont
  {T.}~\bibnamefont {Nakamura}}, \bibinfo {author} {\bibfnamefont
  {H.}~\bibnamefont {Bei}}, \bibinfo {author} {\bibfnamefont {S.}~\bibnamefont
  {Mu}}, \bibinfo {author} {\bibfnamefont {G.~D.}\ \bibnamefont {Samolyuk}},
  \bibinfo {author} {\bibfnamefont {G.~M.}\ \bibnamefont {Stocks}}, \bibinfo
  {author} {\bibfnamefont {J.~A.}\ \bibnamefont {Duffy}}, \bibinfo {author}
  {\bibfnamefont {J.~W.}\ \bibnamefont {Taylor}}, \bibinfo {author}
  {\bibfnamefont {S.~R.}\ \bibnamefont {Giblin}},\ and\ \bibinfo {author}
  {\bibfnamefont {S.~B.}\ \bibnamefont {Dugdale}},\ }\href
  {https://doi.org/10.1103/PhysRevB.102.174405} {\bibfield  {journal} {\bibinfo
   {journal} {Physical Review B}\ }\textbf {\bibinfo {volume} {102}},\ \bibinfo
  {pages} {174405} (\bibinfo {year} {2020})}\BibitemShut {NoStop}%
\bibitem [{\citenamefont {Velişa}\ \emph {et~al.}(2023)\citenamefont
  {Velişa}, \citenamefont {Granberg}, \citenamefont {Levo}, \citenamefont
  {Zhou}, \citenamefont {Fan}, \citenamefont {Bei}, \citenamefont {Tuomisto},
  \citenamefont {Nordlund}, \citenamefont {Djurabekova}, \citenamefont
  {Weber},\ and\ \citenamefont {Zhang}}]{velisa_recent_2023}%
  \BibitemOpen
  \bibfield  {author} {\bibinfo {author} {\bibfnamefont {G.}~\bibnamefont
  {Velişa}}, \bibinfo {author} {\bibfnamefont {F.}~\bibnamefont {Granberg}},
  \bibinfo {author} {\bibfnamefont {E.}~\bibnamefont {Levo}}, \bibinfo {author}
  {\bibfnamefont {Y.}~\bibnamefont {Zhou}}, \bibinfo {author} {\bibfnamefont
  {Z.}~\bibnamefont {Fan}}, \bibinfo {author} {\bibfnamefont {H.}~\bibnamefont
  {Bei}}, \bibinfo {author} {\bibfnamefont {F.}~\bibnamefont {Tuomisto}},
  \bibinfo {author} {\bibfnamefont {K.}~\bibnamefont {Nordlund}}, \bibinfo
  {author} {\bibfnamefont {F.}~\bibnamefont {Djurabekova}}, \bibinfo {author}
  {\bibfnamefont {W.~J.}\ \bibnamefont {Weber}},\ and\ \bibinfo {author}
  {\bibfnamefont {Y.}~\bibnamefont {Zhang}},\ }\href {https://doi.org/10.1557/s43578-023-00922-0} {\bibfield
  {journal} {\bibinfo  {journal} {Journal of Materials Research}\ }\textbf
  {\bibinfo {volume} {38}},\ \bibinfo {pages} {1510} (\bibinfo {year}
  {2023})}\BibitemShut {NoStop}%
\bibitem [{\citenamefont {Tamm}\ \emph {et~al.}(2015)\citenamefont {Tamm},
  \citenamefont {Aabloo}, \citenamefont {Klintenberg}, \citenamefont {Stocks},\
  and\ \citenamefont {Caro}}]{tamm_atomic-scale_2015}%
  \BibitemOpen
  \bibfield  {author} {\bibinfo {author} {\bibfnamefont {A.}~\bibnamefont
  {Tamm}}, \bibinfo {author} {\bibfnamefont {A.}~\bibnamefont {Aabloo}},
  \bibinfo {author} {\bibfnamefont {M.}~\bibnamefont {Klintenberg}}, \bibinfo
  {author} {\bibfnamefont {M.}~\bibnamefont {Stocks}},\ and\ \bibinfo {author}
  {\bibfnamefont {A.}~\bibnamefont {Caro}},\ }\href
  {https://doi.org/10.1016/j.actamat.2015.08.015} {\bibfield  {journal}
  {\bibinfo  {journal} {Acta Materialia}\ }\textbf {\bibinfo {volume} {99}},\
  \bibinfo {pages} {307} (\bibinfo {year} {2015})}\BibitemShut {NoStop}%
\bibitem [{\citenamefont {Ding}\ \emph {et~al.}(2018)\citenamefont {Ding},
  \citenamefont {Yu}, \citenamefont {Asta},\ and\ \citenamefont
  {Ritchie}}]{ding_tunable_2018}%
  \BibitemOpen
  \bibfield  {author} {\bibinfo {author} {\bibfnamefont {J.}~\bibnamefont
  {Ding}}, \bibinfo {author} {\bibfnamefont {Q.}~\bibnamefont {Yu}}, \bibinfo
  {author} {\bibfnamefont {M.}~\bibnamefont {Asta}},\ and\ \bibinfo {author}
  {\bibfnamefont {R.~O.}\ \bibnamefont {Ritchie}},\ }\href
  {https://doi.org/10.1073/pnas.1808660115} {\bibfield  {journal} {\bibinfo
  {journal} {Proceedings of the National Academy of Sciences}\ }\textbf
  {\bibinfo {volume} {115}},\ \bibinfo {pages} {8919} (\bibinfo {year}
  {2018})}\BibitemShut {NoStop}%
\bibitem [{\citenamefont {Pei}\ \emph {et~al.}(2020)\citenamefont {Pei},
  \citenamefont {Li}, \citenamefont {Gao},\ and\ \citenamefont
  {Stocks}}]{pei_statistics_2020}%
  \BibitemOpen
  \bibfield  {author} {\bibinfo {author} {\bibfnamefont {Z.}~\bibnamefont
  {Pei}}, \bibinfo {author} {\bibfnamefont {R.}~\bibnamefont {Li}}, \bibinfo
  {author} {\bibfnamefont {M.~C.}\ \bibnamefont {Gao}},\ and\ \bibinfo {author}
  {\bibfnamefont {G.~M.}\ \bibnamefont {Stocks}},\ }\href
  {https://doi.org/10.1038/s41524-020-00389-1} {\bibfield  {journal} {\bibinfo
  {journal} {npj Computational Materials}\ }\textbf {\bibinfo {volume} {6}},\
  \bibinfo {pages} {122} (\bibinfo {year} {2020})}\BibitemShut {NoStop}%
\bibitem [{\citenamefont {Ghosh}\ \emph {et~al.}(2022)\citenamefont {Ghosh},
  \citenamefont {Sotskov}, \citenamefont {Shapeev}, \citenamefont
  {Neugebauer},\ and\ \citenamefont {Kormann}}]{ghosh_short-range_2022}%
  \BibitemOpen
  \bibfield  {author} {\bibinfo {author} {\bibfnamefont {S.}~\bibnamefont
  {Ghosh}}, \bibinfo {author} {\bibfnamefont {V.}~\bibnamefont {Sotskov}},
  \bibinfo {author} {\bibfnamefont {A.~V.}\ \bibnamefont {Shapeev}}, \bibinfo
  {author} {\bibfnamefont {J.}~\bibnamefont {Neugebauer}},\ and\ \bibinfo
  {author} {\bibfnamefont {F.}~\bibnamefont {Kormann}},\ }\href
  {https://doi.org/10.1103/PhysRevMaterials.6.113804} {\bibfield  {journal}
  {\bibinfo  {journal} {Physical Review Materials}\ }\textbf {\bibinfo {volume}
  {6}},\ \bibinfo {pages} {113804} (\bibinfo {year} {2022})}\BibitemShut {NoStop}%
\bibitem [{\citenamefont {Du}\ \emph {et~al.}(2022)\citenamefont {Du},
  \citenamefont {Yu}, \citenamefont {Shinzato}, \citenamefont {Meng},
  \citenamefont {Sato}, \citenamefont {Li}, \citenamefont {Fan},\ and\
  \citenamefont {Ogata}}]{du_chemical_2022}%
  \BibitemOpen
  \bibfield  {author} {\bibinfo {author} {\bibfnamefont {J.-P.}\ \bibnamefont
  {Du}}, \bibinfo {author} {\bibfnamefont {P.}~\bibnamefont {Yu}}, \bibinfo
  {author} {\bibfnamefont {S.}~\bibnamefont {Shinzato}}, \bibinfo {author}
  {\bibfnamefont {F.-S.}\ \bibnamefont {Meng}}, \bibinfo {author}
  {\bibfnamefont {Y.}~\bibnamefont {Sato}}, \bibinfo {author} {\bibfnamefont
  {Y.}~\bibnamefont {Li}}, \bibinfo {author} {\bibfnamefont {Y.}~\bibnamefont
  {Fan}},\ and\ \bibinfo {author} {\bibfnamefont {S.}~\bibnamefont {Ogata}},\
  }\href {https://doi.org/10.1016/j.actamat.2022.118314} {\bibfield  {journal}
  {\bibinfo  {journal} {Acta Materialia}\ }\textbf {\bibinfo {volume} {240}},\
  \bibinfo {pages} {118314} (\bibinfo {year} {2022})}\BibitemShut {NoStop}%
\bibitem [{\citenamefont {Walsh}\ \emph {et~al.}(2021)\citenamefont {Walsh},
  \citenamefont {Asta},\ and\ \citenamefont
  {Ritchie}}]{walsh_magnetically_2021}%
  \BibitemOpen
  \bibfield  {author} {\bibinfo {author} {\bibfnamefont {F.}~\bibnamefont
  {Walsh}}, \bibinfo {author} {\bibfnamefont {M.}~\bibnamefont {Asta}},\ and\
  \bibinfo {author} {\bibfnamefont {R.~O.}\ \bibnamefont {Ritchie}},\ }\href
  {https://doi.org/10.1073/pnas.2020540118} {\bibfield  {journal} {\bibinfo
  {journal} {Proceedings of the National Academy of Sciences}\ }\textbf
  {\bibinfo {volume} {118}},\ \bibinfo {pages} {e2020540118} (\bibinfo {year}
  {2021})}\BibitemShut {NoStop}%
\bibitem [{\citenamefont {Schonfeld}\ \emph {et~al.}(2019)\citenamefont
  {Schönfeld}, \citenamefont {Sax}, \citenamefont {Zemp}, \citenamefont
  {Engelke}, \citenamefont {Boesecke}, \citenamefont {Kresse}, \citenamefont
  {Boll}, \citenamefont {Al-Kassab}, \citenamefont {Peil},\ and\ \citenamefont
  {Ruban}}]{schonfeld_local_2019}%
  \BibitemOpen
  \bibfield  {author} {\bibinfo {author} {\bibfnamefont {B.}~\bibnamefont
  {Schonfeld}}, \bibinfo {author} {\bibfnamefont {C.~R.}\ \bibnamefont {Sax}},
  \bibinfo {author} {\bibfnamefont {J.}~\bibnamefont {Zemp}}, \bibinfo {author}
  {\bibfnamefont {M.}~\bibnamefont {Engelke}}, \bibinfo {author} {\bibfnamefont
  {P.}~\bibnamefont {Boesecke}}, \bibinfo {author} {\bibfnamefont
  {T.}~\bibnamefont {Kresse}}, \bibinfo {author} {\bibfnamefont
  {T.}~\bibnamefont {Boll}}, \bibinfo {author} {\bibfnamefont {T.}~\bibnamefont
  {Al-Kassab}}, \bibinfo {author} {\bibfnamefont {O.~E.}\ \bibnamefont
  {Peil}},\ and\ \bibinfo {author} {\bibfnamefont {A.~V.}\ \bibnamefont
  {Ruban}},\ }\href {https://doi.org/10.1103/PhysRevB.99.014206} {\bibfield
  {journal} {\bibinfo  {journal} {Physical Review B}\ }\textbf {\bibinfo
  {volume} {99}},\ \bibinfo {pages} {014206} (\bibinfo {year}
  {2019})}\BibitemShut {NoStop}%
\bibitem [{\citenamefont {Niu}\ \emph {et~al.}(2015)\citenamefont {Niu},
  \citenamefont {Zaddach}, \citenamefont {Oni}, \citenamefont {Sang},
  \citenamefont {Hurt}, \citenamefont {LeBeau}, \citenamefont {Koch},\ and\
  \citenamefont {Irving}}]{niu_spin-driven_2015}%
  \BibitemOpen
  \bibfield  {author} {\bibinfo {author} {\bibfnamefont {C.}~\bibnamefont
  {Niu}}, \bibinfo {author} {\bibfnamefont {A.~J.}\ \bibnamefont {Zaddach}},
  \bibinfo {author} {\bibfnamefont {A.~A.}\ \bibnamefont {Oni}}, \bibinfo
  {author} {\bibfnamefont {X.}~\bibnamefont {Sang}}, \bibinfo {author}
  {\bibfnamefont {J.~W.}\ \bibnamefont {Hurt}}, \bibinfo {author}
  {\bibfnamefont {J.~M.}\ \bibnamefont {LeBeau}}, \bibinfo {author}
  {\bibfnamefont {C.~C.}\ \bibnamefont {Koch}},\ and\ \bibinfo {author}
  {\bibfnamefont {D.~L.}\ \bibnamefont {Irving}},\ }\href
  {https://doi.org/10.1063/1.4918996} {\bibfield  {journal} {\bibinfo
  {journal} {Applied Physics Letters}\ }\textbf {\bibinfo {volume} {106}},\
  \bibinfo {pages} {161906} (\bibinfo {year} {2015})}\BibitemShut {NoStop}%
\bibitem [{\citenamefont {Fukushima}\ \emph {et~al.}(2017)\citenamefont
  {Fukushima}, \citenamefont {Katayama-Yoshida}, \citenamefont {Sato},
  \citenamefont {Ogura}, \citenamefont {Zeller},\ and\ \citenamefont
  {Dederichs}}]{fukushima_local_2017}%
  \BibitemOpen
  \bibfield  {author} {\bibinfo {author} {\bibfnamefont {T.}~\bibnamefont
  {Fukushima}}, \bibinfo {author} {\bibfnamefont {H.}~\bibnamefont
  {Katayama-Yoshida}}, \bibinfo {author} {\bibfnamefont {K.}~\bibnamefont
  {Sato}}, \bibinfo {author} {\bibfnamefont {M.}~\bibnamefont {Ogura}},
  \bibinfo {author} {\bibfnamefont {R.}~\bibnamefont {Zeller}},\ and\ \bibinfo
  {author} {\bibfnamefont {P.~H.}\ \bibnamefont {Dederichs}},\ }\href
  {https://doi.org/10.7566/JPSJ.86.114704} {\bibfield  {journal} {\bibinfo
  {journal} {Journal of the Physical Society of Japan}\ }\textbf {\bibinfo
  {volume} {86}},\ \bibinfo {pages} {114704} (\bibinfo {year}
  {2017})}\BibitemShut {NoStop}%
\bibitem [{\citenamefont {Woodgate}\ and\ \citenamefont
  {Staunton}(2022)}]{woodgate_compositional_2022}%
  \BibitemOpen
  \bibfield  {author} {\bibinfo {author} {\bibfnamefont {C.~D.}\ \bibnamefont
  {Woodgate}}\ and\ \bibinfo {author} {\bibfnamefont {J.~B.}\ \bibnamefont
  {Staunton}},\ }\href {https://doi.org/10.1103/PhysRevB.105.115124} {\bibfield
   {journal} {\bibinfo  {journal} {Physical Review B}\ }\textbf {\bibinfo
  {volume} {105}},\ \bibinfo {pages} {115124} (\bibinfo {year}
  {2022})}\BibitemShut {NoStop}%
\bibitem [{\citenamefont {Savrasov}(1996)}]{savrasov_linear-response_1996}%
  \BibitemOpen
  \bibfield  {author} {\bibinfo {author} {\bibfnamefont {S.~Y.}\ \bibnamefont
  {Savrasov}},\ }\href {https://doi.org/10.1103/PhysRevB.54.16470} {\bibfield
  {journal} {\bibinfo  {journal} {Physical Review B}\ }\textbf {\bibinfo
  {volume} {54}},\ \bibinfo {pages} {16470} (\bibinfo {year}
  {1996})}\BibitemShut {NoStop}%
\bibitem [{\citenamefont {Baroni}\ \emph {et~al.}(2001)\citenamefont {Baroni},
  \citenamefont {de~Gironcoli}, \citenamefont {Dal~Corso},\ and\ \citenamefont
  {Giannozzi}}]{baroni_phonons_2001}%
  \BibitemOpen
  \bibfield  {author} {\bibinfo {author} {\bibfnamefont {S.}~\bibnamefont
  {Baroni}}, \bibinfo {author} {\bibfnamefont {S.}~\bibnamefont
  {de~Gironcoli}}, \bibinfo {author} {\bibfnamefont {A.}~\bibnamefont
  {Dal~Corso}},\ and\ \bibinfo {author} {\bibfnamefont {P.}~\bibnamefont
  {Giannozzi}},\ }\href {https://doi.org/10.1103/RevModPhys.73.515} {\bibfield
  {journal} {\bibinfo  {journal} {Reviews of Modern Physics}\ }\textbf
  {\bibinfo {volume} {73}},\ \bibinfo {pages} {515} (\bibinfo {year}
  {2001})}\BibitemShut {NoStop}%
\bibitem [{\citenamefont {Khan}\ \emph {et~al.}(2016)\citenamefont {Khan},
  \citenamefont {Staunton},\ and\ \citenamefont
  {Stocks}}]{khan_statistical_2016}%
  \BibitemOpen
  \bibfield  {author} {\bibinfo {author} {\bibfnamefont {S.~N.}\ \bibnamefont
  {Khan}}, \bibinfo {author} {\bibfnamefont {J.~B.}\ \bibnamefont {Staunton}},\
  and\ \bibinfo {author} {\bibfnamefont {G.~M.}\ \bibnamefont {Stocks}},\
  }\href {https://doi.org/10.1103/PhysRevB.93.054206} {\bibfield  {journal}
  {\bibinfo  {journal} {Phys. Rev. B}\ }\textbf {\bibinfo {volume} {93}},\
  \bibinfo {pages} {054206} (\bibinfo {year} {2016})}\BibitemShut {NoStop}%
\bibitem [{\citenamefont {Woodgate}\ and\ \citenamefont
  {Staunton}(2023)}]{woodgate_short-range_2023}%
  \BibitemOpen
  \bibfield  {author} {\bibinfo {author} {\bibfnamefont {C.~D.}\ \bibnamefont
  {Woodgate}}\ and\ \bibinfo {author} {\bibfnamefont {J.~B.}\ \bibnamefont
  {Staunton}},\ }\href {https://doi.org/10.1103/PhysRevMaterials.7.013801}
  {\bibfield  {journal} {\bibinfo  {journal} {Physical Review Materials}\
  }\textbf {\bibinfo {volume} {7}},\ \bibinfo {pages} {013801} (\bibinfo {year}
  {2023})}\BibitemShut {NoStop}%
\bibitem [{\citenamefont {Gyorffy}\ and\ \citenamefont
  {Stocks}(1983)}]{gyorffy_concentration_1983}%
  \BibitemOpen
  \bibfield  {author} {\bibinfo {author} {\bibfnamefont {B.~L.}\ \bibnamefont
  {Gyorffy}}\ and\ \bibinfo {author} {\bibfnamefont {G.~M.}\ \bibnamefont
  {Stocks}},\ }\href {https://doi.org/10.1103/PhysRevLett.50.374} {\bibfield
  {journal} {\bibinfo  {journal} {Phys. Rev. Lett.}\ }\textbf {\bibinfo
  {volume} {50}},\ \bibinfo {pages} {374} (\bibinfo {year} {1983})}\BibitemShut {NoStop}%
\bibitem [{\citenamefont {Staunton}\ \emph {et~al.}(1994)\citenamefont
  {Staunton}, \citenamefont {Johnson},\ and\ \citenamefont
  {Pinski}}]{staunton_compositional_1994}%
  \BibitemOpen
  \bibfield  {author} {\bibinfo {author} {\bibfnamefont {J.~B.}\ \bibnamefont
  {Staunton}}, \bibinfo {author} {\bibfnamefont {D.~D.}\ \bibnamefont
  {Johnson}},\ and\ \bibinfo {author} {\bibfnamefont {F.~J.}\ \bibnamefont
  {Pinski}},\ }\href {https://doi.org/10.1103/PhysRevB.50.1450} {\bibfield
  {journal} {\bibinfo  {journal} {Physical Review B}\ }\textbf {\bibinfo
  {volume} {50}},\ \bibinfo {pages} {1450} (\bibinfo {year}
  {1994})}\BibitemShut {NoStop}%
\bibitem [{\citenamefont {Khachaturyan}(1978)}]{khachaturyan_ordering_1978}%
  \BibitemOpen
  \bibfield  {author} {\bibinfo {author} {\bibfnamefont {A.~G.}\ \bibnamefont
  {Khachaturyan}},\ }\href {https://doi.org/10.1016/0079-6425(78)90003-8}
  {\bibfield  {journal} {\bibinfo  {journal} {Progress in Materials Science}\
  }\textbf {\bibinfo {volume} {22}},\ \bibinfo {pages} {1} (\bibinfo {year}
  {1978})}\BibitemShut {NoStop}%
\bibitem [{\citenamefont {Bragg}\ and\ \citenamefont
  {Williams}(1934)}]{bragg_effect_1934}%
  \BibitemOpen
  \bibfield  {author} {\bibinfo {author} {\bibfnamefont {W.~L.}\ \bibnamefont
  {Bragg}}\ and\ \bibinfo {author} {\bibfnamefont {E.~J.}\ \bibnamefont
  {Williams}},\ }\href {https://doi.org/10.1098/rspa.1934.0132} {\bibfield
  {journal} {\bibinfo  {journal} {Proceedings of the Royal Society of London.
  Series A, Containing Papers of a Mathematical and Physical Character}\
  }\textbf {\bibinfo {volume} {145}},\ \bibinfo {pages} {699} (\bibinfo {year}
  {1934})}\BibitemShut {NoStop}%
\bibitem [{\citenamefont {Bragg}\ and\ \citenamefont
  {Williams}(1935)}]{bragg_effect_1935}%
  \BibitemOpen
  \bibfield  {author} {\bibinfo {author} {\bibfnamefont {W.~L.}\ \bibnamefont
  {Bragg}}\ and\ \bibinfo {author} {\bibfnamefont {E.~J.}\ \bibnamefont
  {Williams}},\ }\href {https://doi.org/10.1098/rspa.1935.0165} {\bibfield
  {journal} {\bibinfo  {journal} {Proceedings of the Royal Society of London.
  Series A - Mathematical and Physical Sciences}\ }\textbf {\bibinfo {volume}
  {151}},\ \bibinfo {pages} {540} (\bibinfo {year} {1935})}\BibitemShut
  {NoStop}%
\bibitem [{\citenamefont {Mendive-Tapia}\ and\ \citenamefont
  {Staunton}(2017)}]{mendive-tapia_theory_2017}%
  \BibitemOpen
  \bibfield  {author} {\bibinfo {author} {\bibfnamefont {E.}~\bibnamefont
  {Mendive-Tapia}}\ and\ \bibinfo {author} {\bibfnamefont {J.~B.}\ \bibnamefont
  {Staunton}},\ }\href {https://doi.org/10.1103/PhysRevLett.118.197202}
  {\bibfield  {journal} {\bibinfo  {journal} {Physical Review Letters}\
  }\textbf {\bibinfo {volume} {118}},\ \bibinfo {pages} {197202} (\bibinfo
  {year} {2017})}\BibitemShut {NoStop}%
\bibitem [{\citenamefont {Mendive~Tapia}(2020)}]{mendive_tapia_ab_2020}%
  \BibitemOpen
  \bibfield  {author} {\bibinfo {author} {\bibfnamefont {E.}~\bibnamefont
  {Mendive~Tapia}},\ }\href {https://doi.org/10.1007/978-3-030-37238-5} {\emph
  {\bibinfo {title} {Ab initio {Theory} of {Magnetic} {Ordering}: {Electronic}
  {Origin} of {Pair}- and {Multi}-{Spin} {Interactions}}}},\ Springer {Theses}\
  (\bibinfo  {publisher} {Springer International Publishing},\ \bibinfo
  {address} {Cham},\ \bibinfo {year} {2020})\BibitemShut {NoStop}%
\bibitem [{\citenamefont {Oh}\ \emph {et~al.}(2016)\citenamefont {Oh},
  \citenamefont {Ma}, \citenamefont {Leyson}, \citenamefont {Grabowski},
  \citenamefont {Park}, \citenamefont {Körmann},\ and\ \citenamefont
  {Raabe}}]{oh_lattice_2016}%
  \BibitemOpen
  \bibfield  {author} {\bibinfo {author} {\bibfnamefont {H.~S.}\ \bibnamefont
  {Oh}}, \bibinfo {author} {\bibfnamefont {D.}~\bibnamefont {Ma}}, \bibinfo
  {author} {\bibfnamefont {G.~P.}\ \bibnamefont {Leyson}}, \bibinfo {author}
  {\bibfnamefont {B.}~\bibnamefont {Grabowski}}, \bibinfo {author}
  {\bibfnamefont {E.~S.}\ \bibnamefont {Park}}, \bibinfo {author}
  {\bibfnamefont {F.}~\bibnamefont {Körmann}},\ and\ \bibinfo {author}
  {\bibfnamefont {D.}~\bibnamefont {Raabe}},\ }\href
  {https://doi.org/10.3390/e18090321} {\bibfield  {journal} {\bibinfo
  {journal} {Entropy}\ }\textbf {\bibinfo {volume} {18}},\ \bibinfo {pages}
  {321} (\bibinfo {year} {2016})}\BibitemShut {NoStop}%
\bibitem [{\citenamefont {Landau}\ and\ \citenamefont
  {Binder}(2014)}]{landau_guide_2014}%
  \BibitemOpen
  \bibfield  {author} {\bibinfo {author} {\bibfnamefont {D.~P.}\ \bibnamefont
  {Landau}}\ and\ \bibinfo {author} {\bibfnamefont {K.}~\bibnamefont
  {Binder}},\ }\href {https://doi.org/10.1017/CBO9781139696463} {\emph
  {\bibinfo {title} {A {Guide} to {Monte} {Carlo} {Simulations} in
  {Statistical} {Physics}}}},\ \bibinfo {edition} {4th}\ ed.\ (\bibinfo
  {publisher} {Cambridge University Press},\ \bibinfo {address} {Cambridge,
  UK},\ \bibinfo {year} {2014})\BibitemShut {NoStop}%
\bibitem [{\citenamefont {Allen}\ and\ \citenamefont
  {Tildesley}(2017)}]{allen_computer_2017}%
  \BibitemOpen
  \bibfield  {author} {\bibinfo {author} {\bibfnamefont {M.~P.}\ \bibnamefont
  {Allen}}\ and\ \bibinfo {author} {\bibfnamefont {D.~J.}\ \bibnamefont
  {Tildesley}},\ }\href@noop {} {\emph {\bibinfo {title} {Computer Simulation
  of Liquids}}},\ \bibinfo {edition} {2nd}\ ed.\ (\bibinfo
  {publisher} {Oxford University Press},\ \bibinfo {address} {Oxford, United
  Kingdom},\ \bibinfo {year} {2017})\BibitemShut {NoStop}%
\bibitem [{\citenamefont {Cowley}(1950)}]{cowley_approximate_1950}%
  \BibitemOpen
  \bibfield  {author} {\bibinfo {author} {\bibfnamefont {J.~M.}\ \bibnamefont
  {Cowley}},\ }\href {https://doi.org/10.1103/PhysRev.77.669} {\bibfield
  {journal} {\bibinfo  {journal} {Physical Review}\ }\textbf {\bibinfo {volume}
  {77}},\ \bibinfo {pages} {669} (\bibinfo {year} {1950})}\BibitemShut {NoStop}%
\bibitem [{\citenamefont {Cowley}(1965)}]{cowley_short-range_1965}%
  \BibitemOpen
  \bibfield  {author} {\bibinfo {author} {\bibfnamefont {J.~M.}\ \bibnamefont
  {Cowley}},\ }\href {https://doi.org/10.1103/PhysRev.138.A1384} {\bibfield
  {journal} {\bibinfo  {journal} {Physical Review}\ }\textbf {\bibinfo {volume}
  {138}},\ \bibinfo {pages} {A1384} (\bibinfo {year} {1965})}\BibitemShut
  {NoStop}%
\bibitem [{\citenamefont {Martin}(2004)}]{martin_electronic_2004}%
  \BibitemOpen
  \bibfield  {author} {\bibinfo {author} {\bibfnamefont {R.~M.}\ \bibnamefont
  {Martin}},\ }\href@noop {} {\emph {\bibinfo {title} {Electronic {Structure}:
  {Basic} {Theory} and {Practical} {Methods}}}}\ (\bibinfo  {publisher}
  {Cambridge University Press},\ \bibinfo {year} {2004})\BibitemShut {NoStop}%
\bibitem [{\citenamefont {Faulkner}\ and\ \citenamefont
  {Stocks}(1980)}]{faulkner_calculating_1980}%
  \BibitemOpen
  \bibfield  {author} {\bibinfo {author} {\bibfnamefont {J.~S.}\ \bibnamefont
  {Faulkner}}\ and\ \bibinfo {author} {\bibfnamefont {G.~M.}\ \bibnamefont
  {Stocks}},\ }\href {https://doi.org/10.1103/PhysRevB.21.3222} {\bibfield
  {journal} {\bibinfo  {journal} {Physical Review B}\ }\textbf {\bibinfo
  {volume} {21}},\ \bibinfo {pages} {3222} (\bibinfo {year}
  {1980})}\BibitemShut {NoStop}%
\bibitem [{\citenamefont {Faulkner}\ \emph {et~al.}(2018)\citenamefont
  {Faulkner}, \citenamefont {Stocks},\ and\ \citenamefont
  {Wang}}]{faulkner_multiple_2018}%
  \BibitemOpen
  \bibfield  {author} {\bibinfo {author} {\bibfnamefont {J.~S.}\ \bibnamefont
  {Faulkner}}, \bibinfo {author} {\bibfnamefont {G.~M.}\ \bibnamefont
  {Stocks}},\ and\ \bibinfo {author} {\bibfnamefont {Y.}~\bibnamefont {Wang}},\
  }\href {https://doi.org/10.1088/2053-2563/aae7d8} {\emph {\bibinfo {title}
  {Multiple scattering theory electronic structure of solids}}},\ \bibinfo
  {edition} {1st}\ ed.\ (\bibinfo  {publisher} {IOP Publishing},\ \bibinfo
  {address} {Bristol, UK},\ \bibinfo {year} {2018})\BibitemShut {NoStop}%
\bibitem [{\citenamefont {Johnson}\ \emph {et~al.}(1990)\citenamefont
  {Johnson}, \citenamefont {Nicholson}, \citenamefont {Pinski}, \citenamefont
  {Gyorffy},\ and\ \citenamefont {Stocks}}]{johnson_total-energy_1990}%
  \BibitemOpen
  \bibfield  {author} {\bibinfo {author} {\bibfnamefont {D.~D.}\ \bibnamefont
  {Johnson}}, \bibinfo {author} {\bibfnamefont {D.~M.}\ \bibnamefont
  {Nicholson}}, \bibinfo {author} {\bibfnamefont {F.~J.}\ \bibnamefont
  {Pinski}}, \bibinfo {author} {\bibfnamefont {B.~L.}\ \bibnamefont
  {Gyorffy}},\ and\ \bibinfo {author} {\bibfnamefont {G.~M.}\ \bibnamefont
  {Stocks}},\ }\href {https://doi.org/10.1103/PhysRevB.41.9701} {\bibfield
  {journal} {\bibinfo  {journal} {Physical Review B}\ }\textbf {\bibinfo
  {volume} {41}},\ \bibinfo {pages} {9701} (\bibinfo {year}
  {1990})}\BibitemShut {NoStop}%
\bibitem [{\citenamefont {Hoffmann}\ \emph {et~al.}(2020)\citenamefont
  {Hoffmann}, \citenamefont {Ernst}, \citenamefont {Hergert}, \citenamefont
  {Antonov}, \citenamefont {Adeagbo}, \citenamefont {Geilhufe},\ and\
  \citenamefont {Ben~Hamed}}]{hoffmann_magnetic_2020}%
  \BibitemOpen
  \bibfield  {author} {\bibinfo {author} {\bibfnamefont {M.}~\bibnamefont
  {Hoffmann}}, \bibinfo {author} {\bibfnamefont {A.}~\bibnamefont {Ernst}},
  \bibinfo {author} {\bibfnamefont {W.}~\bibnamefont {Hergert}}, \bibinfo
  {author} {\bibfnamefont {V.~N.}\ \bibnamefont {Antonov}}, \bibinfo {author}
  {\bibfnamefont {W.~A.}\ \bibnamefont {Adeagbo}}, \bibinfo {author}
  {\bibfnamefont {R.~M.}\ \bibnamefont {Geilhufe}},\ and\ \bibinfo {author}
  {\bibfnamefont {H.}~\bibnamefont {Ben~Hamed}},\ }\href
  {https://doi.org/10.1002/pssb.201900671} {\bibfield  {journal} {\bibinfo
  {journal} {physica status solidi (b)}\ }\textbf {\bibinfo {volume} {257}},\
  \bibinfo {pages} {1900671} (\bibinfo {year} {2020})}\BibitemShut {NoStop}%
\bibitem [{\citenamefont {Stocks}\ \emph {et~al.}(1978)\citenamefont {Stocks},
  \citenamefont {Temmerman},\ and\ \citenamefont
  {Gyorffy}}]{stocks_complete_1978}%
  \BibitemOpen
  \bibfield  {author} {\bibinfo {author} {\bibfnamefont {G.~M.}\ \bibnamefont
  {Stocks}}, \bibinfo {author} {\bibfnamefont {W.~M.}\ \bibnamefont
  {Temmerman}},\ and\ \bibinfo {author} {\bibfnamefont {B.~L.}\ \bibnamefont
  {Gyorffy}},\ }\href {https://doi.org/10.1103/PhysRevLett.41.339} {\bibfield
  {journal} {\bibinfo  {journal} {Physical Review Letters}\ }\textbf {\bibinfo
  {volume} {41}},\ \bibinfo {pages} {339} (\bibinfo {year} {1978})}\BibitemShut
  {NoStop}%
\bibitem [{\citenamefont {Monkhorst}\ and\ \citenamefont
  {Pack}(1976)}]{monkhorst_special_1976}%
  \BibitemOpen
  \bibfield  {author} {\bibinfo {author} {\bibfnamefont {H.~J.}\ \bibnamefont
  {Monkhorst}}\ and\ \bibinfo {author} {\bibfnamefont {J.~D.}\ \bibnamefont
  {Pack}},\ }\href {https://doi.org/10.1103/PhysRevB.13.5188} {\bibfield
  {journal} {\bibinfo  {journal} {Physical Review B}\ }\textbf {\bibinfo
  {volume} {13}},\ \bibinfo {pages} {5188} (\bibinfo {year}
  {1976})}\BibitemShut {NoStop}%
\bibitem [{\citenamefont {Perdew}\ and\ \citenamefont
  {Wang}(1992)}]{perdew_accurate_1992}%
  \BibitemOpen
  \bibfield  {author} {\bibinfo {author} {\bibfnamefont {J.~P.}\ \bibnamefont
  {Perdew}}\ and\ \bibinfo {author} {\bibfnamefont {Y.}~\bibnamefont {Wang}},\
  }\href {https://doi.org/10.1103/PhysRevB.45.13244} {\bibfield  {journal}
  {\bibinfo  {journal} {Physical Review B}\ }\textbf {\bibinfo {volume} {45}},\
  \bibinfo {pages} {13244} (\bibinfo {year} {1992})}\BibitemShut {NoStop}%
\bibitem [{\citenamefont {Zhang}\ \emph {et~al.}(2015)\citenamefont {Zhang},
  \citenamefont {Stocks}, \citenamefont {Jin}, \citenamefont {Lu},
  \citenamefont {Bei}, \citenamefont {Sales}, \citenamefont {Wang},
  \citenamefont {Béland}, \citenamefont {Stoller}, \citenamefont {Samolyuk},
  \citenamefont {Caro}, \citenamefont {Caro},\ and\ \citenamefont
  {Weber}}]{zhang_influence_2015}%
  \BibitemOpen
  \bibfield  {author} {\bibinfo {author} {\bibfnamefont {Y.}~\bibnamefont
  {Zhang}}, \bibinfo {author} {\bibfnamefont {G.~M.}\ \bibnamefont {Stocks}},
  \bibinfo {author} {\bibfnamefont {K.}~\bibnamefont {Jin}}, \bibinfo {author}
  {\bibfnamefont {C.}~\bibnamefont {Lu}}, \bibinfo {author} {\bibfnamefont
  {H.}~\bibnamefont {Bei}}, \bibinfo {author} {\bibfnamefont {B.~C.}\
  \bibnamefont {Sales}}, \bibinfo {author} {\bibfnamefont {L.}~\bibnamefont
  {Wang}}, \bibinfo {author} {\bibfnamefont {L.~K.}\ \bibnamefont {Béland}},
  \bibinfo {author} {\bibfnamefont {R.~E.}\ \bibnamefont {Stoller}}, \bibinfo
  {author} {\bibfnamefont {G.~D.}\ \bibnamefont {Samolyuk}}, \bibinfo {author}
  {\bibfnamefont {M.}~\bibnamefont {Caro}}, \bibinfo {author} {\bibfnamefont
  {A.}~\bibnamefont {Caro}},\ and\ \bibinfo {author} {\bibfnamefont {W.~J.}\
  \bibnamefont {Weber}},\ }\href {https://doi.org/10.1038/ncomms9736}
  {\bibfield  {journal} {\bibinfo  {journal} {Nature Communications}\ }\textbf
  {\bibinfo {volume} {6}},\ \bibinfo {pages} {8736} (\bibinfo {year}
  {2015})}\BibitemShut {NoStop}%
\bibitem [{\citenamefont {Yin}\ \emph {et~al.}(2020)\citenamefont {Yin},
  \citenamefont {Yoshida}, \citenamefont {Tsuji},\ and\ \citenamefont
  {Curtin}}]{yin_yield_2020}%
  \BibitemOpen
  \bibfield  {author} {\bibinfo {author} {\bibfnamefont {B.}~\bibnamefont
  {Yin}}, \bibinfo {author} {\bibfnamefont {S.}~\bibnamefont {Yoshida}},
  \bibinfo {author} {\bibfnamefont {N.}~\bibnamefont {Tsuji}},\ and\ \bibinfo
  {author} {\bibfnamefont {W.~A.}\ \bibnamefont {Curtin}},\ }\href
  {https://doi.org/10.1038/s41467-020-16083-1} {\bibfield  {journal} {\bibinfo
  {journal} {Nature Communications}\ }\textbf {\bibinfo {volume} {11}},\
  \bibinfo {pages} {2507} (\bibinfo {year} {2020})}\BibitemShut {NoStop}%
\bibitem [{\citenamefont {Gyorffy}\ \emph {et~al.}(1985)\citenamefont
  {Gyorffy}, \citenamefont {Pindor}, \citenamefont {Staunton}, \citenamefont
  {Stocks},\ and\ \citenamefont {Winter}}]{gyorffy_first-principles_1985}%
  \BibitemOpen
  \bibfield  {author} {\bibinfo {author} {\bibfnamefont {B.~L.}\ \bibnamefont
  {Gyorffy}}, \bibinfo {author} {\bibfnamefont {A.~J.}\ \bibnamefont {Pindor}},
  \bibinfo {author} {\bibfnamefont {J.}~\bibnamefont {Staunton}}, \bibinfo
  {author} {\bibfnamefont {G.~M.}\ \bibnamefont {Stocks}},\ and\ \bibinfo
  {author} {\bibfnamefont {H.}~\bibnamefont {Winter}},\ }\href
  {https://doi.org/10.1088/0305-4608/15/6/018} {\bibfield  {journal} {\bibinfo
  {journal} {Journal of Physics F: Metal Physics}\ }\textbf {\bibinfo {volume}
  {15}},\ \bibinfo {pages} {1337} (\bibinfo {year} {1985})}\BibitemShut
  {NoStop}%
\bibitem [{\citenamefont {Rahaman}\ \emph {et~al.}(2014)\citenamefont
  {Rahaman}, \citenamefont {Johansson},\ and\ \citenamefont
  {Ruban}}]{rahaman_first-principles_2014}%
  \BibitemOpen
  \bibfield  {author} {\bibinfo {author} {\bibfnamefont {M.}~\bibnamefont
  {Rahaman}}, \bibinfo {author} {\bibfnamefont {B.}~\bibnamefont {Johansson}},\
  and\ \bibinfo {author} {\bibfnamefont {A.~V.}\ \bibnamefont {Ruban}},\ }\href
  {https://doi.org/10.1103/PhysRevB.89.064103} {\bibfield  {journal} {\bibinfo
  {journal} {Physical Review B}\ }\textbf {\bibinfo {volume} {89}},\ \bibinfo
  {pages} {064103} (\bibinfo {year} {2014})}\BibitemShut {NoStop}%
  \bibitem [{sup()}]{supplementary}%
  \BibitemOpen
  \href@noop {} {}\bibinfo {howpublished}
  {See supplemental material at \url{URL_will_be_inserted_by_publisher} for atom-atom interchange parameters obtained for the alloys studied in this work.}\BibitemShut {NoStop}%
\end{thebibliography}
\end{document}